\documentclass[epj,nopacs]{svjour} 

\usepackage{graphicx}
\usepackage{float}
\usepackage{amsmath}
\usepackage{amssymb}
\usepackage{mathtools}
\usepackage{amsfonts}
\usepackage{dsfont}
\usepackage{nicefrac}
\usepackage{array}
\usepackage{bm}
\usepackage{mathrsfs}
\usepackage{pifont}
\usepackage{multirow}
\usepackage{upgreek}
\usepackage[ruled,lined]{algorithm2e}
\usepackage[dvipsnames]{xcolor}
\usepackage[binary-units=true]{siunitx}
\usepackage[%
  pdftitle={Prony Generalised Eigenvalue Method for Symmetric
    Euclidean Correlation Functions}
  pdfauthor={Johann Ostmeyer and Aniket Sen and Carsten Urbach},
  bookmarks,
  colorlinks,
  linkcolor=myblue,
  citecolor=mymagenta,
  menucolor=black,
  urlcolor=myblue,
  plainpages=false,
  pdfpagelabels,
  hypertexnames=false]{hyperref}
\usepackage{verbatim}
\usepackage{slashed}

\makeatletter
\def\cl@chapter{\@elt {conjecture}}
\makeatother

\usepackage{cleveref}
\usepackage{subfigure}

\usepackage[backend=bibtex,terseinits=false,sorting=none,style=phys,biblabel=brackets,
  defernumbers=true,minnames=15,maxnames=99,
  minsortnames=15, maxsortnames=99,
  minbibnames=15, maxbibnames=99]{biblatex}
\DeclareMathOperator{\ord}{\mathcal{O}}

\DeclareMathOperator{\diag}{diag}

\DeclarePairedDelimiter\abs{|}{|}
\DeclarePairedDelimiter\norm{\lVert}{\rVert}
\newcommand{\eto}[1]{\ensuremath{\mathrm{e}^{#1}}}
\newcommand{\trans}{\ensuremath{\mathsf{T}}}

\newcommand{\ordnung}[1]{\ensuremath{\ord\left(#1\right)}}
\newcommand{\br}[1]{\ensuremath{\left(#1\right)}}
\newcommand{\brr}[1]{\ensuremath{\left[#1\right]}}

\definecolor{mymagenta}{RGB}{200, 0, 100}
\definecolor{myblue}{RGB}{45, 48, 146}

\graphicspath{{plots/},{code/}}

\makeatletter%
\makeatletter%
\begin{document}
\title{The Truncated Hankel Correlator Method}

\author{J.~Ostmeyer\inst{1}
	\and C.~Urbach\inst{1}
}
\authorrunning{J.~Ostmeyer \emph{et al.}}
\titlerunning{Truncated Hankel Correlator Method}
\institute{
  Helmholtz-Institut~für~Strahlen-~und~Kernphysik~and~Bethe~Center~for
  Theoretical~Physics, Universität~Bonn, \\ Bonn,~Germany
}

\date{\today}
\abstract{
	We introduce a new method to approximate Euclidean correlation functions by exponential sums. The Truncated Hankel Correlator (THC) method builds a Hankel matrix from the full correlator data available and truncates the eigenspectrum of said Hankel matrix. It proceeds by applying the Prony generalised eigenvalue method to the thus obtained low-rank approximation. A large number of algebraic correlator analysis methods including (block) Prony (and equivalently (block) Lanczos) and the generalised eigenvalue problem (GEVP) can be reproduced as sub-optimal special cases of the THC method. Weights, for instance inverse square errors, can be included in the analysis, so that the result has a close to optimal $\chi^2$-value. This makes the THC method very similar in spirit to a closed form solution to multi-state fits, naturally including the case of matrix-valued correlators. We show that, in general, finding approximations better than those provided by the THC method is exponentially hard in the number of exponentials. Moreover, the THC method is robust against noise and requires comparably little human oversight. Finally, when applied to symmetric data, the obtained energy spectrum is guaranteed to be symmetric up to machine precision.
}

\maketitle

\section{Introduction}

The spectral analysis of noisy time series plays an
important role in many scientific fields. A particular example is
given by Euclidean correlation functions (imaginary time Matsubara data)
$C(t)$ which result from Monte Carlo simulations of
statistical systems and field theories. $C(t)$ can oftentimes be shown
to be equal to a sum of exponentials in $t$
\begin{equation}
  C(t) = \sum_i c_i \exp\br{-\nicefrac{t}{\tau_i}}
\end{equation}
up to noise inherent to stochastic simulations. Knowledge of the
parameters $c_i$ and $\tau_i$ give access to important properties of
the underlying system.
In lattice quantum chromodynamics (LQCD), for instance, the $c_i$ are
directly related to operator matrix elements, and the $\tau_i$ to
energy eigenvalues $E_i\equiv\nicefrac{1}{\tau_i}$ of the lattice Hamiltonian, which can be related to
hadron masses.

To no surprise, plenty of analysis methods have been developed to
analyse noisy estimators of $C(t)$ for $c_i$ and
$\tau_i$. G.~R.~de~Prony 
went beyond the simple logarithmic derivative of $C(t)$ already in
1795 by inventing a method now named after him, and mapped the problem
to a root finding task. Prony's method has been adapted and applied 
for LQCD in Refs.~\cite{Fleming:2004hs,Beane:2009kya,Cushman:2019hfh,Cushman:2019tcv,Fleming:2023zml}. 

A rather popular\footnote{mainly, because systematics can be estimated
rigorously.} method used in the field of LQCD maps the problem to a
generalised eigenvalue problem
(GEVP)~\cite{Michael:1982gb,Luscher:1990ck,Blossier:2009kd}, i.e. an
algebraic method. In Ref.~\cite{Fischer:2020bgv} it could then be shown
that also Prony's method has an algebraic counterpart by mapping it to
a GEVP based on the Hankel matrix $H$ constructed from $C(t)$ as
\begin{equation}
  H_{ij}(t)\ =\ C(t+i+j)\,.
\end{equation}
As such, it became apparent that the GEVP and the Prony generalised eigenvalue method (PGEVM) are special cases
of the Generalised Pencil of Function (GPoF)~\cite{Aubin:2010jc,Schiel:2015kwa,Ottnad:2017mzd} method.

In Ref.~\cite{Wagman:2024rid} the Lanczos algorithm was used to
determine $\tau_i$ and the author even claimed that the
infamous signal-to-noise (StN) problem was solved by it, which is
defined as the problem of exponential deterioration 
of the StN ratio with increasing Euclidean time~\cite{Lepage:1989}.
However, it was shown in Ref.~\cite{Ostmeyer:2024qgu} that this is
unfortunately not correct.\footnote{We would even claim that it is
  impossible to solve the StN problem at the analysis level without
  further input/assumptions.}
Furthermore, it was proven in Ref.~\cite{Ostmeyer:2024qgu} that Lanczos is mathematically equivalent to a special
case of  the Prony GEVP introduced in
Ref.~\cite{Fischer:2020bgv}. (See also Ref.~\cite{Chakraborty:2024exj}
for an independent re-derivation of the same result.)
Still, Ref.~\cite{Wagman:2024rid} established a link between Krylov
subspace methods and Prony's algebraic method. This was further
related to Ritz values in Ref.~\cite{Abbott:2025yhm}.

It goes far beyond the scope of this work to provide an exhaustive list of methods available for the approximation of time series or functions by exponential sums.
Even the approaches like ours based on a Hankel matrix range from Prony's original method~\cite{Prony:1795} through denoising of the Hankel matrix~\cite{Yu:2024ncm} all the way to finding polynomial roots based on specific eigenvectors of the Hankel matrix~\cite{Beylkin:2005oao}.
While all these methods turn out to be members of the same larger
class, there are also completely unrelated approaches.
For instance, a method based on ordinary differential equations is described in Ref.~\cite{Romiti:2019qim}. 
Another very popular ansatz is the use of multi-state fits (see e.g.\ Refs.~\cite{ExtendedTwistedMass:2021gbo,Rodekamp:2024ixu}) relying on the numerical minimisation of the corresponding $\chi^2$-function
\begin{align}
	\chi^2 &= \sum_t \frac{1}{\sigma_t^2}\br{C(t)-\sum_{i=1}^k c_i \exp\br{-\nicefrac{t}{\tau_i}}}^2
\end{align}
for the parameters $c_i,\tau_i$, given the errors $\sigma_t$.

The fundamental problem in all Prony-related methods is that the noise
will also be described by exponentials, which leaves one with the
challenge to decide which of the exponentials are artefacts, and need
to be removed. The ansatz in the methods above is therefore as follows:
Use the noisy estimator of $C(t)$ as input and extract parameters
$c_i$ and $\tau_i$. Then decide, based on some
criteria, which of these are artefacts of the noise.

In this paper we follow a different approach described in
algorithm~\ref{alg:basic_THC} in the most basic version: we first investigate the
spectrum of the Hankel matrix $H$, truncate the eigenvalues to remove noise from $H$ and only then
determine the spectral parameters from a truncated Hankel matrix algebraically.
We successfully test the method, which we call Truncated Hankel
Correlator (THC) method, for synthetic data as well as real world LQCD
examples.

\begin{algorithm*}[tb]
	\caption{The basic Truncated Hankel Correlator (THC) algorithm, see alg.~\ref{alg:full_THC} for the general version including weights. A ($d$-dimensional matrix-valued) correlator is assumed to have the form $C_{\alpha\beta}(t) = \sum_l c_{l\alpha}c_{l\beta}^\star e^{-E_l t}$. The THC algorithm takes some noisy correlator measurements $C(t)$, $t=0,\dots,T$ (assuming even $T$) and returns the tuple of energies $E_l$, $l=1,\dots,k$ best compatible with the data for the truncation $k\le \left\lfloor\nicefrac{T}{2}\right\rfloor d$. Of course, $k$ does not have to be known a priori and can be looped over or determined dynamically depending on the spectrum. For an open source implementation see Ref.~\cite{hadron:2024}.}\label{alg:basic_THC}
	\SetKwInOut{Input}{input}
	\SetKwInOut{Params}{parameters}
	\SetKwInOut{Output}{output}
	\Params{truncation dimension $k$}
	\Input{correlator measurements $C(0),\dots,C(T)$}
	\Output{energies $E_l$}			
	$\hat{H} \gets
	\begin{pmatrix}
		C(0) & C(1) & \cdots & C(\nicefrac{T}{2}) \\
		C(1) & C(2) & \cdots & C(\nicefrac{T}{2}+1) \!\! \\
		\vdots & \vdots & \ddots & \vdots\\
		C(\nicefrac{T}{2}) & C(\nicefrac{T}{2}+1) & \cdots & C(T) \\
	\end{pmatrix}$ \tcp*{maximal Hankel matrix}
	$U\cdot D\cdot U^\dagger \gets \hat H$ \tcp*{diagonalise, sort $D$ descending by absolute value}
	$(u_1,u_2,\dots) \gets U$ \tcp*{split $U$ into eigenvectors}
	$U_k \gets (u_1,\dots,u_k)$ \tcp*{first $k$ eigenvectors}
	$M_0 \gets P_0 U_k$ \tcp*{projector $P_0= \left(\mathds{1}_{T/2-1},0\right)\otimes \mathds{1}_d$ removes the last time slice}
	$M_1 \gets P_1 U_k$ \tcp*{projector $P_1= \left(0,\mathds{1}_{T/2-1}\right)\otimes \mathds{1}_d$ removes the first time slice}
	\uIf(\tcp*[f]{symmetric correlator}){$C(t) = C(T-t)$}{$\bar M \gets \frac12 \br{M_0+M_1}$\;
	$X \gets \br{\bar M^\dagger M_0}^{-1} \bar M^\dagger M_1$}
	\Else{$X \gets \br{M_0^\dagger M_0}^{-1} M_0^\dagger M_1$\;}
	$V \cdot \diag\br{\Lambda_1,\dots,\Lambda_k}\cdot V^{-1} \gets X$ \tcp*{diagonalise $X$ (not symmetric)}
	$E_l \gets -\log \Lambda_l$\;
\end{algorithm*}

Based on the results of Refs.~\cite{Beylkin:2005oao,ottaviani2014exact}, we show that the THC method provides the optimal solution to a very close approximation of the $\chi^2$-function and that no method with guaranteed polynomial runtime can optimise the exact $\chi^2$-function.
Moreover, it requires one to chose
the truncation dimension as the only parameter of the algorithm, which
can be inferred from the correlator itself with relatively little effort.
The THC method also provides a very natural way to incorporate certain
symmetries of $C(t)$ in Euclidean time $t$.

The rest of the paper is structured as follows.
In \cref{sec:theory} we derive the THC method and collect all the details in the \cref{alg:full_THC,alg:coefficients}.
We also prove the optimality of the THC method and show that virtually all known algebraic correlator analysis methods are special cases of the THC algorithm class in \cref{sec:weights,sec:relations}, respectively.
Practical guidelines for the THC usage are provided in \cref{sec:practice} and numerical experiments are presented in \cref{sec:results}.

\section{Derivation and theory}\label{sec:theory}

Prony's original method~\cite{Prony:1795} relies on the determination of roots of a polynomial and has long since been replaced by an equivalent GEVP that is numerically much more stable~\cite{Fischer:2020bgv}. This algebraic approach has the additional advantage that it can be readily generalised to matrix-valued correlator functions~\cite{Ostmeyer:2024qgu}.
The algorithm derived in the following has been inspired by Ref.~\cite{Zhang:2024mpr} and is the algebraic counterpart to the algorithm introduced in Ref.~\cite{Beylkin:2005oao} and neatly summarised in Ref.~\cite{Beylkin:2010abe}.

Beylkin and Monzón showed in Ref.~\cite{Beylkin:2005oao} that the best approximation to a tower of exponentials with a reduced number of terms can be constructed in the following way. Build the full Hankel matrix and diagonalise it (or, more generally, use a singular value decomposition (SVD)). Choose an eigenvector $u_k$ with the corresponding eigenvalue $|s_k|<\varepsilon$ with some previously defined precision threshold $\varepsilon$. Determine the roots of the polynomial who's coefficients are defined by $u_k$. These roots correspond to the exponentials $\eto{-E_l}$ of the desired energy levels $E_l$. If $s_k=0$, then this algorithm reproduces Prony's original method and extracts all contributing exponentials exactly. As we show below, similarly to Prony's original method, the algorithm by Beylkin and Monzón has an algebraic counterpart that does not require polynomial root finding and can be generalised to matrix-valued correlators. It relies on the same idea that the Hankel matrix is diagonalised and only the most significant $k$ eigenvalues are kept for the analysis. The standard PGEVM is restored when $k$ is large enough to include all non-zero eigenvalues.

\subsection{The Hankel matrix}

Let us start with a hermitian $d\times d$ correlator matrix
\begin{align}
	C_{\alpha\beta}(t) &= \sum_l c_{\alpha,l}c_{\beta,l}^\star e^{-E_l t}\,,\label{eq:corr_matrix}\\
	c_{\alpha,l} &= \langle 0|O_\alpha|l\rangle
\end{align}
for some set of operators $O_\alpha$.
The goal of the following derivation is to reconstruct the energies $E_l$ and coefficients $c_{\alpha,l}$ from the measured (noisy) time series $C(t)$.

Consider an integer tuple $(n,m)$ so that $n+m=T$ and the corresponding maximal rectangular Hankel matrix
\begin{align}
	\hat H^{nm}_{ij} &\coloneqq C(i+j)\,,\ 0\le i\le n\,,\ 0\le j\le m\,, \text{ i.e.} \\
	\hat H^{nm} &=  
	\begin{pmatrix}
		C(0) & C(1) & \cdots & C(m-1) & C(m) \\
		C(1) & C(2) & \cdots & C(m) & C(m+1) \!\! \\
		\vdots & \vdots & \ddots & \vdots & \vdots\\
		\!\! C(n-1) & C(n) & \cdots & C(T-2) & C(T-1) \\
		C(n) & C(n+1) & \cdots & C(T-1) & C(T) \\
	\end{pmatrix}\,,\label{eq:maximal_rect_hankel}
\end{align}
where each correlator entry $C(t)$ can itself be a square matrix of dimension $d\ge 1$. Assuming even $T$, we can define the special case
\begin{align}
	\begin{split}		
		\hat{H} &\coloneqq \hat H^{\nicefrac{T}{2},\nicefrac{T}{2}}\\
		&=
		\begin{pmatrix}
			C(0) & C(1) & \cdots & C(\nicefrac{T}{2}-1) & C(\nicefrac{T}{2}) \\
			C(1) & C(2) & \cdots & C(\nicefrac{T}{2}) & C(\nicefrac{T}{2}+1) \!\! \\
			\vdots & \vdots & \ddots & \vdots & \vdots\\
			\!\! C(\nicefrac{T}{2}-1) & C(\nicefrac{T}{2}) & \cdots & C(T-2) & C(T-1) \\
			C(\nicefrac{T}{2}) & C(\nicefrac{T}{2}+1) & \cdots & C(T-1) & C(T) \\
		\end{pmatrix}\,.
	\end{split}\label{eq:maximal_hankel}
\end{align}
For $T$ odd, choose $\hat H$ as the largest possible square matrix (i.e.\ ignore the last time slice $C(T)$).
Unless stated otherwise explicitly, we will stick to the square matrix $\hat H$ throughout the rest of this manuscript because it has the highest possible rank and thus allows the extraction of the highest amount of information.

Note that $\hat H$ inherits the hermiticity from $C(t)$ since
\begin{align}
	\hat H_{ij}^\dagger &= C(i+j)^\dagger\\
	&= C(i+j) = C(j+i)\\
	&= \hat H_{ji}\,.
\end{align}

Every ``classical'' Hankel matrix
\begin{equation}
	H_{ij}(t) = C(t + i\Delta + j\Delta)\,,\ 0\leq i,j < n\,,
	\label{eq:hankel}
\end{equation}
can be constructed as a submatrix from $\hat H$ by choosing $n$
equidistant rows and columns with integer distance $\Delta$, starting with the $t$-th row. The equidistance, however, is not strictly necessary. Any square submatrix of $\hat H$ can be used to construct a GEVP of Prony type~\cite{Fischer:2020bgv}
\begin{equation}
	H(t+\delta t)\cdot v_l(t)\ =\ \Lambda_l^n(t, \delta t) H(t)\cdot v_l(t)\,.
	\label{eq:gevp}
\end{equation}
In \cref{sec:block_Prony_proof} we provide a proof that such a PGEVM leads to the correct energies and coefficients, even if the correlator $C(t)$ is matrix-valued.

\subsection{The truncation}

From this point on we will replace the element-wise notation by a matrix-product based one. Every choice of indices discussed above can be re-written as a projection onto a given sub-space. In particular, we define the projector
\begin{align}
	P^{(n)}_\delta &\coloneqq \left(\underbrace{0,\dots,0}_{\delta \text{times}},\mathds{1}_n,0,\dots\right)\otimes \mathds{1}_d \label{eq:projector}
\end{align}
onto an $nd$-dimensional subspace starting after the $\delta$-th row, where $d$ is the correlator matrix dimension. Moreover, we introduce the general $m\times(nd)$ weight matrices $W$, $W'$ and the invertible $\br{\br{T/2+1}d}\times \br{\br{T/2+1}d}$ (i.e.\ maximally dimensional) weights $\Omega$. Then, for a given shift $\delta t$, a further generalised GEVP defined by the Prony formalism reads
\begin{align}
	W\, P_{\delta t} \, \hat H {P_0}^\trans {W'}^\dagger \cdot v_l &= \Lambda_l\, W\, P_{0} \, \hat H {P_0}^\trans {W'}^\dagger \cdot v_l\,,\label{eq:full_GEVP_matrix-form}
\end{align}
where the projector superscripts $n=\frac T2 -\delta t$ are maximal
and have been dropped. The form~\eqref{eq:hankel} for $H$ is regained for weights
\begin{align}
	W_{ij} = 
	W'_{ij} = \begin{cases}
		1 & \text{if } i = j = 0\pmod \Delta\,,\\
		0 & \text{else.}
	\end{cases}
\end{align}

To date, the usual approach to stabilising this GEVP in the presence
of noise has been to vary $\Delta$, $\delta t$ and $n$
(or equivalently weights $W,W'$) while keeping the Hankel matrix $\hat H$ exact. We now propose to choose the weights in accordance with the inverse statistical uncertainties and to ensure stability by approximating $\hat H$. For this we define and diagonalise the auxiliary matrix
\begin{align}
	\tilde H &\coloneqq \Omega \hat H \Omega^\dagger\label{eq:weighted_full_Hankel}\\
	&= U D U^\dagger\,,
\end{align}
where $U=\br{u_1,u_2,\dots}$ is a unitary matrix (i.e.\ $UU^\dagger=\mathds{1}$) with eigenvectors $u_i$ and $D=\mathrm{diag}\br{s_1,s_2,\dots}$ is diagonal with all $s_i\in\mathds{R}$ and $|s_1|\ge |s_2|\ge \cdots$. This diagonalisation is possible because $\hat H$ and, in consequence, $\tilde H$ are hermitian (typically real and symmetric). Now, for any given subspace dimension $k$ the best possible low-rank matrix approximation of $\tilde H$ (in Frobenius norm) is given by
\begin{align}
	\tilde H \approx \tilde{H}_k &\coloneqq U_k D_k U_k^\dagger\,,\label{eq:low-rank_approximation}\\
	D_k &\coloneqq \mathrm{diag}\br{s_1,\dots,s_k}\,,\\
	U_k &\coloneqq \br{u_1,\dots,u_k}\,.
\end{align}
Therefore the best $\Omega$-weighted approximation of the full Hankel matrix is (using the notation $\Omega^{-\dagger}\equiv\br{\Omega^{-1}}^{\dagger}=\br{\Omega^{\dagger}}^{-1}$)
\begin{align}
	\hat H &\approx \hat H_k \coloneqq \Omega^{-1} \tilde H_k \Omega^{-\dagger}\,.\label{eq:omega-approx}
\end{align}
The subspace dimension $k$ should be chosen so that the physical states can be resolved accurately while noise is removed. Since all physical states should come with positive eigenvalues (and the weights $\Omega$ preserve positive definiteness), good results can be achieved by taking $k$ to be the largest value so that all considered eigenvalues $s_i>0$, $i\le k$ are positive.

The approximation~\eqref{eq:omega-approx} can be inserted into the GEVP~\eqref{eq:full_GEVP_matrix-form}
\begin{align}
	W\, P_{\delta t} \, \Omega^{-1} U_k \cdot \tilde v_l &= \tilde\Lambda_l\, W\, P_{0} \, \Omega^{-1} U_k \cdot \tilde v_l\,,\label{eq:approx_GEVP}\\
	\tilde v_l &\coloneqq D_k U_k^\dagger\Omega^{-\dagger} {P_0}^\trans {W'}^\dagger \cdot v_l\,,
\end{align}
where the generalised eigenvalues $\tilde \Lambda_l$ are guaranteed to approximate the $k$ dominant true eigenvalues $\Lambda_l$. At this point, we have an overdetermined GEVP with up to $\frac T2 - 1$ equations but only $k$ eigenvalues and eigenvectors. In general, such a problem does not have an exact solution. However, there is a unique solution that minimises the residual error. To extract it, we first rewrite equation~\eqref{eq:approx_GEVP} as a system of linear equations
\begin{align}
	M_{\delta t} &= M_0 \cdot X\,,\\
	M_\delta &\coloneqq W\, P_{\delta} \, \Omega^{-1} U_k\,,\\
	X &\coloneqq \br{\tilde v_1,\dots,\tilde v_k} \cdot \mathrm{diag}\br{\tilde\Lambda_1,\dots,\tilde\Lambda_k} \cdot \br{\tilde v_1,\dots,\tilde v_k}^{-1}\,.
\end{align}
The residuum $||M_{\delta t}-M_0 \cdot X||_F$ in Frobenius norm is minimised by
\begin{align}
	X &= \br{M_0^\dagger M_0}^{-1} M_0^\dagger M_{\delta t}\,.\label{eq:get_X_stable}
\end{align}
Finally, diagonalising $X$ yields the desired generalised eigenvalues $\tilde\Lambda_l\approx \eto{-E_l \delta t}$.

Since the difference in the norm effectively occurs at the times of $M_{\delta t}$, it makes sense to choose the weights $W$ with the same shift
\begin{align}
		W &= P_{\delta t} \hat W {P_{\delta t}}^\trans\,,\label{eq:asym_weights}\\		 
		\begin{split}
			\hat W &\coloneqq \mathrm{diag}\left(\frac{1}{\sqrt{\sigma_{11,0}}}, \dots, \frac{1}{\sqrt{\sigma_{dd,0}}},\right.\\
			&\qquad \frac{1}{\sqrt{\sigma_{11,2}}}, \dots, \frac{1}{\sqrt{\sigma_{11,4}}}, \dots\\
			&\qquad \left.\frac{1}{\sqrt{\sigma_{dd,T-2}}}, \dots, \frac{1}{\sqrt{\sigma_{dd,T}}}\right)\,,
		\end{split}\label{eq:diag_weights}
\end{align}
where $\sigma_{\alpha\beta,t}$ denotes the error of $C_{\alpha\beta}(t)$, to obtain a good approximation of the $\chi^2$-function.

For time-symmetric correlators, we advocate to use the symmetrised version
\begin{align}
	X &= \br{\bar M^\dagger M_0}^{-1} \bar M^\dagger M_{\delta t}\,,\label{eq:get_X_sym}\\
	\bar M &\coloneqq \frac12 \br{M_0+M_{\delta t}}\,,\\
	W &= \sqrt{P_{0} \hat W^2 {P_0}^\trans+P_{\delta t} \hat W^2 {P_{\delta t}}^\trans}\label{eq:sym_weights}
\end{align}
which equally considers the residuum $||M_{\delta t}\cdot X^{-1}-M_0||_F$.

Note that the symmetrised version using $\bar M$ guarantees an exact symmetry between $\tilde \Lambda_l$ and $\tilde \Lambda_l^{-1}$ for symmetric correlators. This means that the spectrum of energies $E_l$ will be symmetric about zero up to machine precision. The inversion of $M_0^\dagger M_0$ is, however, numerically more stable and should therefore be used for non-symmetric correlators.

Note that in practice the very first time slices can be entirely dominated by excited states and thus useless for the analysis aiming for low energies. Therefore, it might be beneficial to slightly reduce $\hat H$, starting from $t_0> 0$. This should be the first thing to try if an out-of-the box analysis as described above does not produce satisfactory results.

The full Truncated Hankel Correlator (THC) algorithm (including good default choices) is summarised in algorithm~\ref{alg:full_THC}.

\begin{algorithm*}[tb]
	\caption{The full Truncated Hankel Correlator (THC) algorithm, see alg.~\ref{alg:basic_THC} for a simplified version. A ($d$-dimensional matrix-valued) correlator is assumed to have the form $C_{\alpha\beta}(t) = \sum_l c_{\alpha,l}c_{\beta,l}^\star e^{-E_l t}$. The THC algorithm takes some noisy correlator measurements $C(t)$, $t=0,\dots,T$ (assuming even $T$) and returns the tuple of energies $\tilde E_l$, $l=1,\dots,k$ best compatible with the data for the truncation $k\le \left\lfloor\nicefrac{T}{2}\right\rfloor d$ given the weights $W$,$\Omega$. Of course, $k$ does not have to be known a priori and can be looped over or determined dynamically depending on the spectrum. For an open source implementation see Ref.~\cite{hadron:2024}.}\label{alg:full_THC}
	\SetKwInOut{Input}{input}
	\SetKwInOut{Params}{parameters}
	\SetKwInOut{Output}{output}
	\Params{truncation dimension $k$, shift $\delta t$ (default $\delta t=1$)}
	\Input{correlator measurements $C(0),\dots,C(T)$ with uncertainties $\sigma\brr{C(0)},\dots,\sigma\brr{C(T)}$,\\
	weight matrices $W$,$\Omega$ (defaults in eqs.~(\ref{eq:asym_weights},\ref{eq:diag_weights},\ref{eq:sym_weights},\ref{eq:diag_weight_mat}))}
	\Output{energies $\tilde E_l$}			
	$\hat{H} \gets
	\begin{pmatrix}
		C(0) & C(1) & \cdots & C(\nicefrac{T}{2}) \\
		C(1) & C(2) & \cdots & C(\nicefrac{T}{2}+1) \!\! \\
		\vdots & \vdots & \ddots & \vdots\\
		C(\nicefrac{T}{2}) & C(\nicefrac{T}{2}+1) & \cdots & C(T) \\
	\end{pmatrix}$ \tcp*{maximal Hankel matrix}
	$\tilde H \gets \Omega \hat H \Omega^\dagger$\;
	$(u_1,u_2,\dots)\cdot D\cdot (u_1,u_2,\dots)^\dagger \gets \tilde H$ \tcp*{diagonalise, sort $D$ descending by absolute value}
	$U_k \gets (u_1,\dots,u_k)$ \tcp*{first $k$ eigenvectors}
	$M_0 \gets W P_0 \Omega^{-1} U_k$ \tcp*{projector $P_0\equiv P_0^{(\nicefrac{T}{2}-\delta t)}$ from eq.~\eqref{eq:projector}}
	$M_{\delta t} \gets W P_{\delta t} \Omega^{-1} U_k$ \tcp*{projector $P_{\delta t}\equiv P_{\delta t}^{(\nicefrac{T}{2}-\delta t)}$ from eq.~\eqref{eq:projector}}
	\uIf(\tcp*[f]{symmetric correlator}){$C(t) = C(T-t)$}{$\bar M \gets \frac12 \br{M_0+M_1}$\;
		$X \gets \br{\bar M^\dagger M_0}^{-1} \bar M^\dagger M_1$}
	\Else{$X \gets \br{M_0^\dagger M_0}^{-1} M_0^\dagger M_1$\;}
	$\tilde V \cdot \diag\br{\tilde\Lambda_1,\dots,\tilde\Lambda_k}\cdot \tilde V^{-1} \gets X$ \tcp*{diagonalise}
	$\tilde E_l \gets -\frac{1}{\delta t}\log \tilde\Lambda_l$\;
\end{algorithm*}

\subsection{Reconstructing coefficients}

The coefficients $c_{\alpha,l}$ in equation~\eqref{eq:corr_matrix} can be reconstructed by inverting the Vandermonde matrix
\begin{align}
	\chi_{tl} &= \eto{-E_l t}
\end{align}
or solving the minimal residual problem again. For this, we interpret each time series of correlator matrix elements $C_{\alpha\beta}(t)$ as a vector in $t$ and the unknown matrix element estimators
\begin{align}
	\tilde c_{\alpha\beta,l} &\equiv \tilde c_{\alpha,l}\tilde c_{\beta,l}^\star\label{eq:coeff_matrix}
\end{align}
as a vector in $l$. This allows us to relate these quantities in a compact matrix notation
\begin{align}
	\chi \cdot \tilde c_{\alpha\beta} &= C_{\alpha\beta}
\end{align}
for any given $(\alpha,\beta)$. Again, we are faced with an
overdetermined system of linear equations. As before, the residual
deviation can be minimised exactly by
\begin{align}
	\tilde c_{\alpha\beta} &= \br{\chi^\dagger \sigma_{\alpha\beta}^{-2} \chi}^{-1} \chi^\dagger \sigma_{\alpha\beta}^{-2} C_{\alpha\beta}\,,\label{eq:get_weighted_coeffs}
\end{align}
including the weight matrix $\sigma_{\alpha\beta}^{-2}$ built from the inverse uncertainties of the respective correlator measurements.

Such a reconstruction of the coefficients $\tilde c_{\alpha\beta,l}$ does not include the theoretical constraint~\eqref{eq:coeff_matrix}. This property is only restored in the limit of large $k$ or, more accurately, when the approximation $\tilde H_k=\tilde H$ becomes exact. Thus, in practice another decomposition is required when extracting the coefficients $\tilde c_{\alpha,l}$.
Note that such a decomposition is only reliable for physical states (e.g.\ in LQCD the vector form $c_{\alpha,l}$ corresponds to the matrix elements) that guarantee the structure of equation~\eqref{eq:corr_matrix}. Noisy states come without any intrinsic structure so that, in general, the matrix $\tilde c_{\alpha\beta,l}$ will not be close to rank-1, or even positive definite.

In realistic scenarios the Vandermonde matrix spans many orders of magnitude (e.g.\ over $T=40$ time slices an energy of $E= \pm1$ contributes values between $\eto{ET}\approx\num{2e17}$ and $\eto{-ET}\approx\num{4e-18}$). Thus, for a stable implementation, it is crucial to rescale the Vandermonde matrix in such a way that it becomes well-conditioned and accurately invertible. A good solution is to split $\chi$ into
\begin{align}
	\chi &= \chi^0 \chi^D\,,\\
	\chi^D &\coloneqq \diag\br{\max_t\left|\eto{-E_l t}\right|}\,,
\end{align}
so that all elements $\left|\chi^0_{tl}\right|\le 1$ are bounded and the uniform bound $\max_t \left|\chi^0_{tl}\right|= 1$ is reached for every $l$. The diagonal and thus trivially invertible matrix $\chi^D$ sets the respective scale.

Using this decomposition of the Vandermonde matrix and inserting it into equation~\eqref{eq:get_weighted_coeffs}, finally yields the numerically viable procedure
\begin{align}	
	\tilde c_{\alpha\beta} &= \br{\chi^D}^{-1}\br{{\chi^0}^\dagger \sigma_{\alpha\beta}^{-2} \chi^0}^{-1} {\chi^0}^\dagger \sigma_{\alpha\beta}^{-2} C_{\alpha\beta}\,,\\
	\tilde c_{\alpha\beta,l} &\approx \tilde c_{\alpha,l} {\tilde c_{\beta,l}}^{\star}
\end{align}
for the extraction of the coefficients. The decomposition of the matrix defined by $\tilde c_{\alpha\beta,l}$ into the vector form is either again achieved via diagonalisation, dropping all but the maximal eigenvalue, or by using the square roots of the diagonal $|\tilde c_{\alpha,l}|=\sqrt{\tilde c_{\alpha\alpha,l}}$. We advocate the latter method because it is more robust when the coefficients $\tilde c_{\alpha\beta,l}$ have greatly varying scales.

\begin{algorithm*}[tb]
	\caption{The THC coefficient (matrix element) reconstruction, to be used with energies obtained from alg.~\ref{alg:basic_THC} or alg.~\ref{alg:full_THC}. A ($d$-dimensional matrix-valued) correlator is assumed to have the form $C_{\alpha\beta}(t) = \sum_l c_{\alpha,l}c_{\beta,l}^\star e^{-E_l t}$. This second part of the THC algorithm takes some noisy correlator measurements $C(t)$, $t=0,\dots,T$ and a tuple of energies $E_l$, $l=1,\dots,k$ and returns the coefficients $\tilde c_{\alpha,l}$ best compatible with the data and the energies $E_l$ given the uncertainties $\sigma$. Loops over open indices $\alpha,\beta,l$ and $t$ are implicit. For an open source implementation see Ref.~\cite{hadron:2024}.}\label{alg:coefficients}
	\SetKwInOut{Input}{input}
	\SetKwInOut{Params}{parameters}
	\SetKwInOut{Output}{output}
	\Input{correlator measurement vector $C=\br{C(0),\dots,C(T)}$ with uncertainties $\sigma$,
		energies $E_l$}
	\Output{coefficient vectors $\tilde c_{\alpha,l}$ and matrices $\tilde c_{\alpha\beta,l} \approx \tilde c_{\alpha,l} {\tilde c_{\beta,l}}^{\star}$}			
	$\chi_{tl} \gets \eto{-E_l t}$ \tcp*{Vandermonde matrix}
	$\chi^{-D} \gets \diag\br{\max_t\left|\chi_{tl}\right|}^{-1}$ \tcp*{split scales for numerical stability}
	$\chi^0 \gets \chi\cdot \chi^{-D}$\;
	$\tilde c_{\alpha\beta} \gets \chi^{-D} \br{{\chi^0}^\dagger \sigma_{\alpha\beta}^{-2} \chi^0}^{-1} {\chi^0}^\dagger \sigma_{\alpha\beta}^{-2} C_{\alpha\beta}$\;
	$\tilde c_{\alpha,l} \gets \sqrt{\tilde c_{\alpha\alpha,l}}\,\frac{\tilde c_{\alpha1,l}}{|\tilde c_{\alpha1,l}|}$ \tcp*{main contribution vector}
\end{algorithm*}

\subsection{Including weights}\label{sec:weights}

The main difference between the problem solved by Beylkin and Monzón and that realistically encountered in data analysis is the presence of uncertainties. An optimal description of the data should take these uncertainties into account. Presume we want to approximate the measured correlator values $C_{\alpha\beta}(t)$ with errors $\sigma_{\alpha\beta,t}$, or more generally the covariance matrix $\Sigma$, using a function $C_{\alpha\beta}^{(k)}(t)$ with a number of free parameters limited by $k$. Then the best approximation (under the assumption of normally-distributed errors) minimises the corresponding $\chi^2$-function
\begin{align}
	\chi^2_\sigma &= \sum_{\alpha,\beta,t} \abs*{\frac{C_{\alpha\beta}(t)-C_{\alpha\beta}^{(k)}(t)}{\sigma_{\alpha\beta,t}}}^2\,, \text{ or} \label{eq:chi_sq}\\
	\chi^2_\Sigma &= \br{C_{\alpha\beta}(t)-C_{\alpha\beta}^{(k)}(t)}^{\!\!*}\!\!\! \br{\Sigma^{-1}}^{\alpha\beta,t}_{\alpha'\beta',t'}\! \br{C_{\alpha'\beta'}(t')-C_{\alpha'\beta'}^{(k)}(t')}
\end{align}
with the summations over all repeated indices implicit in the last formula.
Thus, the approximation $\hat H_k$ of the Hankel matrix $\hat H$ should minimise residuals $\chi^2_{\sigma,\Sigma}$ and not simply the Frobenius norm as we have done so far using an SVD. More rigorously, we are interested in the \textit{weighted} low-rank approximation~\cite{Srebro:2003wlra} $\hat H_k$ that fulfils
\begin{align}
	\hat H_k &= \min_{H} \sum_{\alpha,\beta,t} \frac{\abs*{\hat H_{tt',\alpha\beta}-H_{tt',\alpha\beta}}^2}{\sigma_{\alpha\beta,t+t'}^2\, \#(t+t')}\,, \text{ or}\\
	\hat H_k &= \min_{H} \mathrm{vec}\br{\hat H-H}^\dagger \Sigma_{\#}^{-1}\, \mathrm{vec}\br{\hat H-H}\,,
\end{align}
subject to $\mathrm{rank}\br{H}\le k$, respectively. Here, we introduced the multiplicity counter
\begin{align}
	\#(t) &= \br{\frac T2+1-\left|\frac T2 - t\right|}\label{eq:multiplicity}
\end{align}
of a given time in order to counteract the disproportionate influence of repeated entries in the Hankel matrix. This multiplicity is already included in the effective covariance matrix $\Sigma_{\#}$ acting on the vectorised matrix $\mathrm{vec}\br{\hat H-H}$.

Our approximation~\eqref{eq:low-rank_approximation} based on a truncation of the eigenvalues allows non-uniform weights introduced through the matrix $\Omega$. The specific choice
\begin{align}
	\begin{split}
		\Omega &= \mathrm{diag}\left(\frac{1}{\sqrt{\sigma_{11,0}}}, \dots, \frac{1}{\sqrt{\sigma_{dd,0}}},\right.\\
		&\qquad \frac{1}{\sqrt{\sqrt{3}\sigma_{11,2}}}, \dots, \frac{1}{\sqrt{\sqrt{5}\sigma_{11,4}}}, \dots\\
		&\qquad \left.\frac{1}{\sqrt{\sqrt{3}\sigma_{dd,T-2}}}, \dots, \frac{1}{\sqrt{\sigma_{dd,T}}}\right)\,,
	\end{split}\label{eq:diag_weight_mat}
\end{align}
reproduces the optimal weights of the uncorrelated $\chi^2$-function on the diagonal of the Hankel matrix
\begin{align}
	\tilde H_{tt,\alpha\alpha} &= \frac{\hat{H}_{tt,\alpha\alpha}}{\sigma_{\alpha\alpha,2t}\, \sqrt{\#(2t)}}\,.
\end{align}
Iterative procedures to obtain better weights are derived and discussed in Ref.~\cite{Zvonarev:2017}, but equation~\eqref{eq:diag_weight_mat} is likely as close as we can get to the fully weighted case without a drastic increase in computational complexity.

A simple way to put these mathematical considerations into perspective is the following: the THC method provides the exact solution to an approximation of the problem, while methods like multi-state fits provide approximate solutions to the exact problem.

\subsection{Optimality and computational complexity}

Unfortunately, finding weighted low-rank approximations is NP-hard~\cite{Gillis:2011lrma}, even in the uncorrelated case. In particular, no closed-form solution exists, as opposed to the SVD in the unweighted case. Here, we are faced with the slightly simpler case of a Hankel structured low-rank approximation which is not NP-hard in the dimension of the Hankel matrix~\cite{gillard:2023}. The problem is, however, still exponentially hard in the truncation (or rank) $k$. More specifically, the computational complexity grows at least with the number of local minima $\ordnung{T^k}$ of the corresponding $\chi^2$-function~\cite{ottaviani2014exact}. Therefore, in general it is infeasible to find the truly optimal matrix $H_k$. Since this algebraic formulation is exactly equivalent to the initial problem of finding the optimal tower of exponentials approximating the measured correlators, we conclude that the initial problem is exponentially hard in the rank $k$ as well. This is the very reason why multi-state fits become highly unstable for $k\gtrsim 4$ states. In summary, the pursuit of a closed-form solution (based on an SVD or otherwise) to this fitting problem is doomed to fail.

That said, good (if not provably optimal) weighted low-rank approximations can be found numerically. It is not unlikely that at least for weights with small fluctuations the algorithms presented in Ref.~\cite{Srebro:2003wlra} or similar methods can yield the optimal approximation for a given rank $k$ within the runtime of just a few diagonalisations. An additional problem then is to identify the optimal rank $k$. In the unweighted case, $k$ can be estimated from the spectrum of eigenvalues. In the weighted case, however, the familiar ``greedy'' method no longer works. That is, the space spanning the optimal rank $k$ approximation is not necessarily a subspace of the best rank $k+1$ approximation.

Based on these considerations, we can rigorously pin down in which sense the THC method is optimal, which class of problems would in principle yield even better results and which subset of the latter class has no guaranteed polynomial runtime solutions. For this, we will quote the following theorems adjusted for our notation. They are not new and their proofs are provided in the references quoted, respectively.

\begin{theorem}[Ref.~\cite{gillard:2023}, Theorem~2.2 with $Q=R=\Omega^\dagger \Omega$]
	Given the matrix-valued correlator data $C(t)$, the weight matrix $\Omega$ and the truncation $k$, then the approximation $\hat H_k$ as in eq.~\eqref{eq:omega-approx} has minimal residuum
	\begin{align}
		\hat H_k &= \min_{H:\mathrm{rank}\br{H} \le k}\, \norm*{\Omega\br{\hat H-H}\Omega^\dagger}_F
	\end{align}
	with $\hat H$ as in eq.~\eqref{eq:maximal_hankel}.
\end{theorem}

\begin{remark}
	This means that the THC algorithm~\ref{alg:full_THC} uses the best possible approximation of the data $C(t)$ with $k$ states, i.e.\ it features the fastest possible convergence, under the $\Omega$-weighted Frobenius norm.
\end{remark}

\begin{corollary}[Ref.~\cite{gillard:2023}, eq.~(6)]
	Given the matrix-valued correlator data $C(t)$, the weight matrix $\Omega$ and the error tolerance $\varepsilon$, then the minimal truncation
	\begin{align}
		k &= \min\left\{j\,|\ \mathrm{rank}\br{H} \le j,\ \norm*{\Omega\br{\hat H-H}\Omega^\dagger}_F < \varepsilon\right\}
	\end{align}
	satisfying the tolerance requirements also suffices for
	\begin{align}
		\norm*{\Omega\br{\hat H-H_k}\Omega^\dagger}_F < \varepsilon
	\end{align}
	with $\hat H$ as in eq.~\eqref{eq:maximal_hankel} and $\hat H_k$ as in eq.~\eqref{eq:omega-approx}.
\end{corollary}

\begin{remark}
	This dual formulation means that the THC algorithm~\ref{alg:full_THC} uses the smallest possible number of states $k$ to approximate the data $C(t)$ with $\varepsilon$-accuracy under the $\Omega$-weighted Frobenius norm. The THC method's high robustness against noise is a direct consequence. Another implication arises for any method using only a part of the data and a lower rank $k'$ (e.g.\ the effective mass with only two time slices and $k'=1$). The results have to be compatible with those from THC on the full data for said $k'$ states, or otherwise they will have contaminations larger than $\varepsilon$ from unresolved states. Thus, it is always optimal (under the $\Omega$-weighted Frobenius norm) to use the full data with the THC method.
\end{remark}

\begin{theorem}[Ref.~\cite{ottaviani2014exact}, Corollary~3.9]
	Given the scalar correlator data $C(t)$ with corresponding uncertainties $\sigma_{t}$, the $\chi^2$-function~\eqref{eq:chi_sq} with the approximation
	\begin{align}
		C^{(k)}(t) &= \sum_{l=1}^k c_l \eto{-E_l t}
	\end{align}
	has exponentially many local minima in the number $2k$ of free parameters $c_l,E_l$.
\end{theorem}

\begin{remark}
	This means that the best approximation of the data $C(t)$ by $k$ states with respect to the true $\chi^2$-function cannot be found in polynomial time in general. Since this result holds for scalar correlators, it is certainly true for correlator matrices. It is worth noting that even the simple case of constant $\sigma_t=1$ does not have a closed-form solution~\cite{gillard:2023,Zvonarev:2017}.
\end{remark}

\subsection{Counting the degrees of freedom}

It should not come as a surprise that, independently of the (invertible) weights $\Omega$ and $W$, the THC method reproduces the exact sum of exponentials once the truncation dimension $k$ is at least equal to the number of non-zero exponential contributions. This is a direct consequence of the Hankel matrix becoming singular once all states can be resolved. Then the truncation of zero eigenvalues does not change the matrix.

A more interesting question is, therefore, how many states $k$ can be resolved, given some number of time slices $T+1$ (i.e.\ $t=0,\dots,T$) and correlator matrix dimension $d$. A na\"ive approach might simply count the degrees of freedom (dof) generated by the $d^2$ matrix elements per time slice. This would overestimate $k$ in general because the Hankel matrix is required to be symmetric (or hermitian) and the matrix elements are thus not all independent.
Instead, every symmetric time slice contributes $\frac{d(d+1)}{2}$ dof. At the same time, $k$ states require $k$ exponents (energies) and $dk$ coefficients. Put together, the total number of dof reads
\begin{align}
	\#\text{dof} &= (T+1)\frac{d(d+1)}{2} - (d+1)k\,.
\end{align}
The number of dof cannot be negative, dictating
\begin{align}
	k \le k_\text{sup} &\coloneqq \left\lfloor\frac{(T+1)d}{2}\right\rfloor\,.
\end{align}

It is reassuring that the maximal dimension of the projected Hankel matrix
\begin{align}
	k_\text{max} \coloneqq \dim\br{P_0\hat H P_0^\trans} &= \left\lfloor\frac{T}{2}\right\rfloor d
\end{align}
does not exceed the supremum $k_\text{max}\le k_\text{sup}$, demonstrating that the THC method is always well-defined.
Somewhat surprisingly, for many tuples of $T$ and $d$, the supremum is not reached by the maximum. More specifically
\begin{align}
	k_\text{max} &= \begin{cases}
		 k_\text{sup} - d & \text{if $T$ odd,}\\
		 k_\text{sup} -\frac d2 & \text{if $T$,$d$ even,} \\
		 k_\text{sup} - \frac{d-1}{2} & \text{if $T$ even, $d$ odd.}
	\end{cases}
\end{align}
Overall, the number of states that cannot be resolved compared to the total information hidden in the data, that is the relative information loss, is of order $\ordnung{\frac 1T}$. In most realistic scenarios this loss is negligible.

Even this small discrepancy can be lifted in the odd $T$ case (e.g.\ $T=1$, $d=1$ produce the effective mass). While most likely not useful in practice, $\hat H$ can be replaced by $\hat H^{\br{\nicefrac{(T+1)}{2},\nicefrac{(T-1)}{2}}}$, simply adding a row to $\hat H$, see eq.~\eqref{eq:maximal_rect_hankel}. Then the diagonalisation during the THC process has to be replaced by an SVD.
For even $T$, on the other hand, improvement beyond $k=k_\text{max}$ is much more complicated and not always possible. For instance, $T=0$ leaves the exponents undetermined for any $d$. We will not try to salvage the few potentially remaining states since typically $T\gg1$ in realistic scenarios and the desired truncation is very small $k\ll k_\text{max}$ anyway.

\subsection{Relations to other methods}\label{sec:relations}

The THC algorithm as described above (using the full square matrix $\hat H$ and monitoring convergence in the truncation $k$) has the highest information gain and convergence rate among a vast class of THC-type methods. Put differently, $\hat H$ captures the largest possible Krylov subspace of the Euclidean time evolution operator in the full (infinite-dimensional) Hilbert space. The truncated matrix $\hat H_k$ used for the THC analysis is the best possible low-rank approximation (up to exponentially hard variations using weights) of this Krylov subspace. This means that for any given $k$ the least amount of information is lost and for any given precision requirement the smallest possible $k$ suffices. The latter property is particularly helpful to make the algorithm robust in the presence of noise.

Below, we will outline a few well-known approaches that turn out to be special THC cases.

By construction the THC method becomes equivalent to the standard (block) Prony or Lanczos algorithms~\cite{Ostmeyer:2024qgu} in the limit of maximal truncation dimension (i.e.\ no truncation at all) and a weight matrix $W=\diag\br{1,\dots,1,0,\dots,0}$ with equal non-zero weights in the first $n$ elements. Clearly, neither of these changes can have any positive effect on the analysis. At best, these established methods are not significantly worse than the THC algorithm introduced here.

Some alternatives to our truncation approach have also been considered before. For instance, Ref.~\cite{Yu:2024ncm} introduces a denoising procedure during which the Hankel matrix is projected onto the closest (in Frobenius norm) positive semi-definite Hankel matrix. This can be done iteratively by alternately removing all negative eigenvalues and averaging back to Hankel form. Convergence is guaranteed because both Hankel matrices and positive semi-definite matrices form compact spaces. For high-dimensional noisy systems this procedure ends up with rather high residual rank matrices with most modes still originating from noise. After all, having a negative eigenvalue is a sufficient but by no means necessary criterion for a noise mode (see fig.~\ref{fig:synthetic_eigenvalues} where for synthetic data the 7th to 12th positive eigenvalue modes are known to be noise). It is tempting from these considerations to denoise the Hankel matrix by projecting it to the closest Hankel matrix of rank $k$ (instead of the closest arbitrary matrix of rank $k$ as in the THC approach). This, however, turns out to be a much harder optimisation problem since rank-$k$-matrices do not form a convex space. This potentially NP-hard overhead is clearly not worth the effort, especially, since a priori it is not even clear that restoring Hankel form will be beneficial.

We note that the column-type matrix $\hat H^{T0}$ immediately yields the standard GEVP or, for $d=1$, the effective mass. Interestingly (if not really useful), for $d>1$ the THC method using an SVD allows to generalise the GEVP via possible truncation. The full energy level extraction including the plateau fit after the GEVP solution can be reproduced by the residuum minimisation provided in equation~\eqref{eq:get_X_stable}.

In principle the row-type matrix $\hat H^{0T}$ can be analysed in the same way. This would effectively amount to fitting $d$ states to the correlator first and subsequently extracting the energies.

Both extreme cases capture much less information than the square matrix based THC algorithm since their rank and thus the maximal number of resolvable states is limited by $d$.

\section{Practical considerations}\label{sec:practice}

The THC method as summarised in algorithm~\ref{alg:full_THC} considerably reduces the amount of human oversight required to obtain reliable results compared to most other methods. For instance, it completely avoids the usual search for a plateau characteristic for effective mass-type fits. However, even with the help of THC, we could not derive a fail-safe fully automatised procedure and some sanity checks or even tuning remain necessary.

\subsection{Choosing the truncation $k$}

Most prominently, the truncation $k$ has to be chosen akin to the number of states in a multi-state fit. This choice comes with the usual trade-off. A systematic error is induced in case $k$ is too small to fully capture the physically relevant features of the correlator. On the other hand, very large $k$ include unnecessarily many noisy states and typically result in avoidably large statistical errors. As a rule, it is safer to choose the truncation $k$ slightly too large rather than too small.

The initial method proposed by Beylkin and Monzón~\cite{Beylkin:2005oao} relied on fixing some threshold $\varepsilon$ upfront and discarding all eigenvalues $s_i$ of the Hankel matrix $\tilde H$ below that threshold $|s_i|<\varepsilon$. While this procedure sounds intuitive, we found that it rarely works with noisy data. The choice of $\varepsilon$ is at least as difficult and ambiguous as that of the truncation $k$ itself which is why we will not consider it any further.

In an ideal scenario, there is a clear scale separation between physically relevant states and noise. In this case, the eigenvalues $s_i$ exhibit a large gap by absolute value. This motivates the choice
\begin{align}
	k_\text{gap} &= \arg\min_i \left|\frac{s_i}{s_{i+1}}\right|
\end{align}
as a good truncation candidate for correlators with relatively little noise. We find that for noisy correlators typically $k_\text{gap}<k_\text{opt}$ underestimates the optimal truncation $k_\text{opt}$.

Another natural candidate for $k$ comes from the theoretical insight that the Hankel matrix constructed from physical correlators must be positive-definite. Thus, it makes sense to truncate right before the first negative (more accurately non-positive) eigenvalue of $\tilde H$
\begin{align}
	k_\text{pos} &= \max\left\{k\,|\, \forall i\le k:s_i>0\right\}
\end{align}
to obtain the largest-rank positive-definite approximation. Interestingly, in practice $k_\text{pos}\approx k_\text{opt}$ is often close to the optimal truncation but prone to very large statistical errors so that $k_\text{pos}\pm 1$ tend to be better choices.

Finally, the safest option (still involving human intervention) to extract the truncation $k$ is to identify a `plateau' and choose the first $k$ compatible with all the later ones. Such a `plateau' shows that the choice does not have a systematic error from too small $k$. Choosing the beginning of the `plateau' minimises the contamination by noise.

\subsection{Adjusting the weights $\Omega$}

Technically speaking, the THC algorithm~\ref{alg:full_THC} is guaranteed to converge for every choice of invertible weight matrices $\Omega$.
The convergence speed and robustness against noise, however, can vary greatly depending on the choice of $\Omega$.
These weights can influence the relative significance of the different states and determine which states are kept at a given level of truncation.
In contrast, the `outer' weights $W$ are usually far less important (as long as they do not contain too many zero eigenvalues).

Our theoretically motivated default weights~\eqref{eq:diag_weight_mat} lead to good results for many but not all examples we have tested.
In particular, systems with a very high noise level tend to converge better with constant weights $\Omega=\mathds{1}$.
On the other hand, one can easily imagine cases where the early and late times are virtually useless due to contaminations from excited states.
In such a scenario it might be beneficial to suppress these contributions stronger than just with their inverse statistical errors.

Should the THC results not appear reasonable with the default weights~\eqref{eq:diag_weight_mat}, we therefore recommend to try out constant weights $\Omega=\mathds{1}$ instead.

\subsection{Filtering for physical states}

In the analysis of physical data, it is very common to be exclusively interested in real energies $\tilde E_l\in \mathds{R}$. For instance, the exact (noiseless) Euclidean time correlator can only contain decay modes and no oscillations. In consequence, all energies $\tilde E_l\not\in \mathds{R}$ with non-vanishing imaginary part are caused by noise and can safely be discarded as unphysical. This first filtering step is well-known and part of all similar methods like Lanczos~\cite{Wagman:2024rid} and Prony~\cite{Ostmeyer:2024qgu}.

The ground state energy, i.e.\ the smallest positive $E_l$, is of particular interest. In our experience, the ground state energy is usually very robust to extract. The only adjustment we find necessary, is the introduction of some small $\epsilon>0$. Then we identify the ground state energy as $\min_l \tilde E_l$ subject to $\tilde E_l>\epsilon$. This shift allows to get rid of spurious energies compatible with zero. The explicit choice of $\epsilon$ does not influence the THC result at all, as long as $\epsilon$ is much smaller than the expected ground state energy and much larger than machine precision.

Especially for symmetric correlators this is very different from the standard Prony method where the ground state can be significantly contaminated by the back-propagating states. The inbuilt symmetry of the THC method, on the other hand, does not allow for energies below that of the ground state other than energies compatible with 0 up to machine precision.

Let us also remark that no filtering by coefficient like the Cullum-Willoughby (CW)~\cite{CULLUM1981329} test or similar is required. Instead, all filtering by significance is already taken care of by the truncation. This makes the THC algorithm much more robust and easier to automatise than Lanczos and similar methods.

\subsection{Interpreting the results}

A successful application of the THC algorithm will produce compatible energy estimators for all truncations $k\ge k_0$ starting from some minimal required truncation $k_0$. This behaviour is visually reminiscent of an effective mass plateau, but it has an entirely different meaning. In fact, the better analogy is that every estimator stems from a $k$-state fit to the same data. Thus, the statistical information is identical for all estimators and all deviations have systematic origin, i.e.\ the deviations are model-dependent. A stable `plateau' now indicates that the model dependence and in consequence the systematic bias from a given choice is negligible. Conversely, the absence of such a `plateau' implies that there is some pathological problem exceeding statistical fluctuations that needs further investigation.

Since the THC method always acts on the full data set, every single point in the `plateau' can be used as a representative for further analysis. Typically, lower truncation orders $k$ lead to numerically more stable and less noisy results. Thus, as a rule, one should stick to points earlier in the constant region so as to not overestimate the uncertainty. This observation also motivates our rather low candidates for optimal truncations $k$.

The `plateau' from a THC analysis should not be fitted since all data
points are maximally correlated.

\section{Numerical Experiments}\label{sec:results}

\subsection{Synthetic Data}

We first look at an artificially generated noiseless correlation function (same as in Ref.~\cite{Ostmeyer:2024qgu})
\begin{equation}
	\label{eq:syn}
	C(t) = \sum_{l=0}^{N_s-1}\ e^{-E_l t}
\end{equation}
with $N_s=6$ and
\begin{equation}
	E_{0, 1, 2, 3, 4, 5} = \{0.06, 0.1, 0.13, 0.18, 0.22, 0.25\}\,.
\end{equation}

\begin{figure}
	\centering
	\includegraphics[width=\linewidth,page=1]{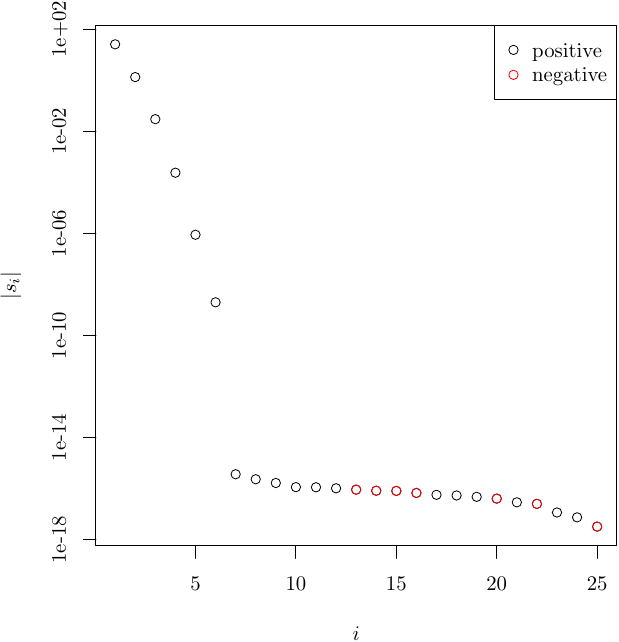}
	\caption{Modulus of the eigenvalues $s_i$ of the full Hankel matrix $\hat H$ defined by eq.~\eqref{eq:maximal_hankel} built from the synthetic correlator~\eqref{eq:syn} with $T=48$. Negative eigenvalues are marked red.}
	\label{fig:synthetic_eigenvalues}
\end{figure}

\begin{figure*}
	\centering
	\includegraphics[width=.45\linewidth,page=2]{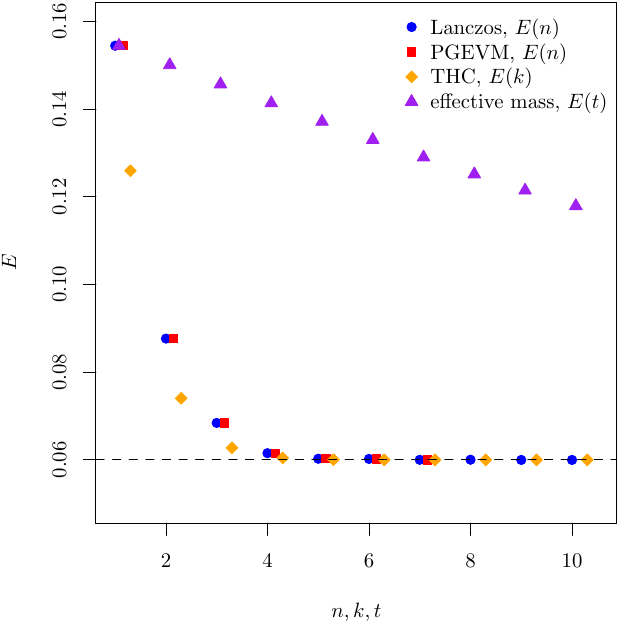}
	\hfill
	\includegraphics[width=.45\linewidth,page=3]{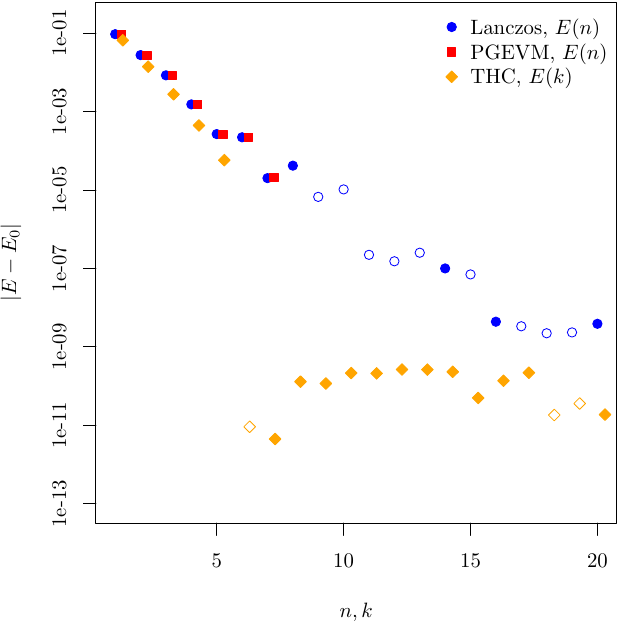}
	\caption{Comparison of Lanczos, PGEVM with $\delta t=1, \Delta=1, t_0=0$, THC using eq.~\eqref{eq:get_X_stable} with uniform weights and $T=48$,
		and effective mass for artificial data, see Eq.~\ref{eq:syn}. All but the Lanczos data is shifted slightly in
		$x$-direction for legibility. Left: convergence of the ground
		state energy level as a function of $n$, $k$, or $t$, respectively. The exact value is
		indicated by the dashed line. Right: the
		difference to the exact ground state energy is plotted on a
		log-scale as a function of $n$ or $k$.
		Empty symbols in the right panel indicate negative differences.}
	\label{fig:synthetic_convergene}
\end{figure*}

As expected, the full Hankel matrix $\hat H$ has exactly $N_s$
non-vanishing eigenvalues $s_i>0$ (all others are zero within machine precision). The corresponding spectrum is visualised in figure~\ref{fig:synthetic_eigenvalues}. Correspondingly, the THC algorithm~\ref{alg:basic_THC} with any truncation $k\ge N_s$ reproduces the exact energies $E_l$. In particular, the convergence of the ground state energy in $k$ is shown in figure~\ref{fig:synthetic_convergene} and compared to that of Lanczos/PGEVM in the Hankel dimension $n$ and to the effective mass in the time $t$. It does not come as a surprise that the convergence of the THC method in $k$ is not only the fastest, it also saturates instantly at a level of machine precision for $k=N_s$.

Since no $k<N_s$ can possibly resolve all the states, the convergence of the THC method to the full sum of exponentials is optimal. This optimality is preserved for arbitrary weight matrices $W$,$\Omega$ in the general THC algorithm~\ref{alg:full_THC}. The deviations of the ground state energy from the exact value as a function of $k<N_s$, on the other hand, depend on the weights. For instance, larger weights towards later times would accelerate the convergence of the ground state even further. While not the main reason, this is certainly an additional motivation of including non-trivial $W$,$\Omega$ into the analysis of noisy data.

Next, we add a back-propagating part
\begin{equation}
	\label{eq:syn_sym}
	C_\text{sym}(t) = 2\sum_{l=0}^{N_s-1} c_l \cosh\brr{E_l \br{t-\frac T2}}
\end{equation}
with $N_s=6$ and
\begin{align}
	c_{0, 1, 2, 3, 4, 5} &= \{1, 0.5, 0.1, 0.05, 0.01, 0.005\}\\
	E_{0, 1, 2, 3, 4, 5} &= \{0.06, 0.1, 0.13, 0.18, 0.22, 0.25\}\,.
\end{align}

\begin{figure}
	\centering
	\includegraphics[width=\linewidth,page=4]{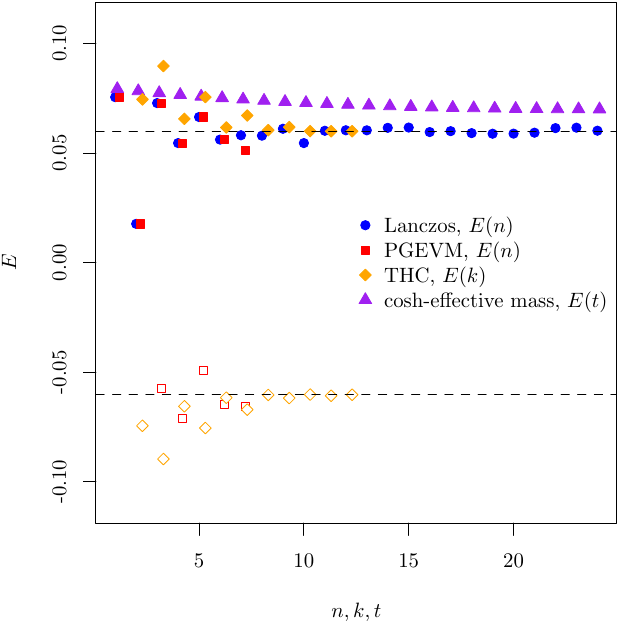}
	\caption{Comparison of Lanczos, PGEVM with $\delta t=1, \Delta=1, t_0=0$, and THC using eq.~\eqref{eq:get_X_sym} with uniform weights and $T=48$,
		and cosh-effective mass
		for symmetric artificial data, see Eq.~\ref{eq:syn_sym}. All but the Lanczos data is shifted slightly in
		$x$-direction for legibility. Convergence of the ground
		state energy level as a function of $n$, $k$, or $t$, respectively. The exact values (positive and negative) are
		indicated by the dashed lines.
		Empty symbols indicate negative values.}
	\label{fig:synthetic_sym}
\end{figure}

Figure~\ref{fig:synthetic_sym} shows the convergence to the ground state in this case. The cosh-effective mass $m_\mathrm{cosh}$ is the numerical solution to the equation
\begin{align}
	\frac{C(t)}{C(t+1)} &= \frac{\cosh\brr{m_\mathrm{cosh} \br{t-\frac T2}}}{\cosh\brr{m_\mathrm{cosh} \br{t+1-\frac T2}}}
\end{align}
and has been included for comparison instead of the standard effective mass. In this case, a truncation of $k=2N_s=12$ is needed to resolve all the positive and negative energies exactly. Once more, this truncation is indeed sufficient for the THC method to reproduce the exact tower of states, that is the THC convergence is again optimal.

While it is harder to tell in this case whether the ground state energy is approached faster by the THC or the Lanczos algorithm, the former has some major advantages. First of all, the inbuilt symmetry of equation~\eqref{eq:syn_sym} used for the THC analysis here leads to an energy spectrum that is exactly symmetric about 0. Consequently, no cross-contamination or misclassification is possible. Therefore, the (positive) ground state is approached strictly from above by the THC estimators allowing an interpretation of $E(k)$ as a rigorous upper bound for every $k$.

The alternating convergence pattern of the THC algorithm is easy to understand es well. The symmetry of the energy spectrum dictates the existence of an exact $E=0$ energy state for odd truncations $k$. To compensate for this contribution, all the other energies are shifted to slightly larger absolute values. If such a state at $E=0$ can be excluded a priori, we therefore advocate to use only even values of $k$.

Truncations $k>2N_s$ require inversions to (numerically) singular matrices which is why no THC points are plotted beyond $k=2N_s$. This was to be expected since $k=N_s$ already allows to resolve all the states exactly. In a way, it is more surprising that this does not happen in the purely decaying case plotted in figure~\ref{fig:synthetic_convergene}. Note also that the PGEVM becomes unstable at an even earlier point while the Lanczos method continues to fluctuate about the true value instead of converging exactly.

\subsection{LQCD Sample Data}

Next we apply the THC method to Euclidean correlation functions
estimated in Monte Carlo simulations for lattice quantum
chromodynamics (LQCD). From now on all energy levels and times are in
lattice units if not stated otherwise. Thus, $E\,\cdot t$ is
dimensionless, while $t$ has the dimension of the lattice spacing $a$ and $E$ the
dimension of $1/a$. 

The statistical uncertainties in this subsection have been determined
using the bootstrap procedure. For pion and $\omega$ meson we apply
\emph{outlier removal} combined with bias correction as discussed in
Ref.~\cite{Ostmeyer:2024qgu}. 
For the nucleon case we apply double bootstrap as discussed in
Refs.~\cite{Ostmeyer:2024qgu,Wagman:2024rid}.

\subsubsection{Pion Correlator}

We start with the pion, which represents the least
challenging case in LQCD, because it can be shown that the
signal-to-noise ratio is independent of Euclidean time~\cite{Lepage:1989}.
We study a correlator matrix of size $2\times2$
\begin{equation}
  C^\pi_{\alpha\beta}(t-t') = \langle O_\alpha^\dagger(t)\, O_\beta(t')\rangle\,,
\end{equation}
with $\alpha,\beta\in\{1,2\}$. The first operator is given as
$O_1 = i\bar\psi\gamma_5\psi$, while $O_2$ is a smeared
version of $O_1$. For the smearing we use fuzzing as described in
detail in Refs.~\cite{ETM:2008zte,Lacock:1994qx}.

In more detail, the LQCD ensemble was generated with
$N_f=2+1+1$ Wilson twisted mass quarks and is denoted as the $B55.32$
ensemble in Ref.~\cite{Ottnad:2012fv} with $L = 32$ and $T = 64$. It has a pion mass
value of about 300 MeV at a lattice spacing of $a =
0.0779(4)\ \mathrm{fm}$.
The correlator estimate is based on $4996$
independent measurements. As a reference value for the ground state energy
level we use $E_\pi = 0.15566(12)$ determined from a fit to the
$\cosh$ effective mass of the $C^\pi_{11}$ element in the time range from $14$ to $27$ with the
error statistical only. Depending on the fit range the reference value
can change by two $\sigma$. This reference value is in good agreement
with values published in the literature, see
Refs.~\cite{Ottnad:2017bjt,Ottnad:2012fv,Baron:2010bv}. 

Due to translational invariance, $C^\pi_{\alpha\beta}$ only depends on the time
difference $t-t'$. Moreover, due to periodic boundary conditions and
the pion quantum numbers, the correlator matrix is real, symmetric in $\alpha,\beta$
and symmetric in time $C^\pi_{\alpha\beta}(t) = C^\pi_{\alpha\beta}(T-t)$. 

\begin{figure}
	\centering
	\includegraphics[width=\linewidth,page=3]{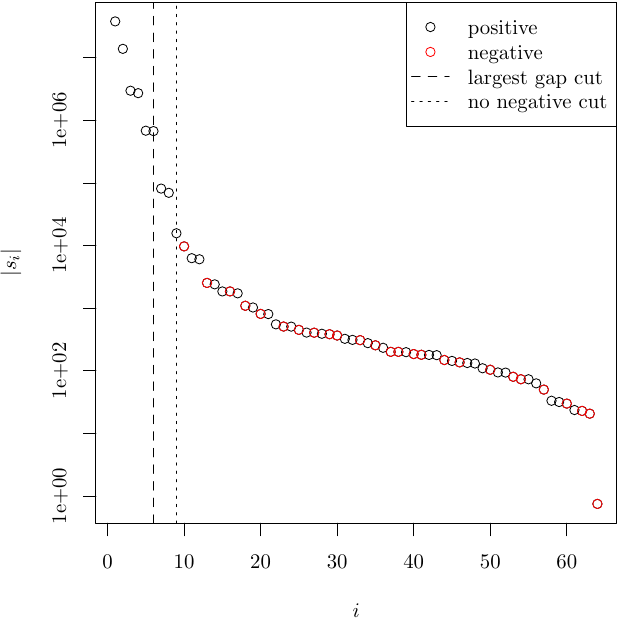}
	\caption{Modulus of the eigenvalues $s_i$ of the full Hankel matrix $\tilde H$ defined by eq.~\eqref{eq:weighted_full_Hankel} built from the pion correlator starting at $t_0=1$ and using the default weights~\eqref{eq:diag_weight_mat}. Negative eigenvalues are marked red. Natural candidates for the truncation $k$ are $k=6$ at the largest ratio $\left|\frac{s_{k}}{s_{k+1}}\right|$ (i.e.\ the largest gap on the log-scale) and $k=9$ right before the first negative eigenvalue $s_{k+1}\le 0$.}
	\label{fig:pion_eigenvalues}
\end{figure}

In figure~\ref{fig:pion_eigenvalues} we plot the modulus of the
eigenvalues $s_i$ of the full Hankel matrix $\tilde H$ with default
weights eq.~\eqref{eq:diag_weight_mat} constructed from this pion
correlator data as a function of their index $i$ sorted decreasing in
modulus.
Negative eigenvalues are
represented by red symbols. By the vertical lines we indicate the two
cut criteria: the first negative eigenvalue is the tenth largest. The
corresponding cut would mean to retain only the $k=9$ largest
eigenvalues. The largest logarithmic gap occurs between sixth and
seventh eigenvalue, and only the $k=6$ largest would be kept in the
reconstruction. These two cuts correspond to removing about $0.1\%$
and $0.4\%$ of the trace of $\tilde H$, respectively, i.e. for $k=6$
the retained eigenvalues account for more than $99\%$ of the trace.

\begin{figure}
	\centering
	\includegraphics[width=\linewidth,page=6]{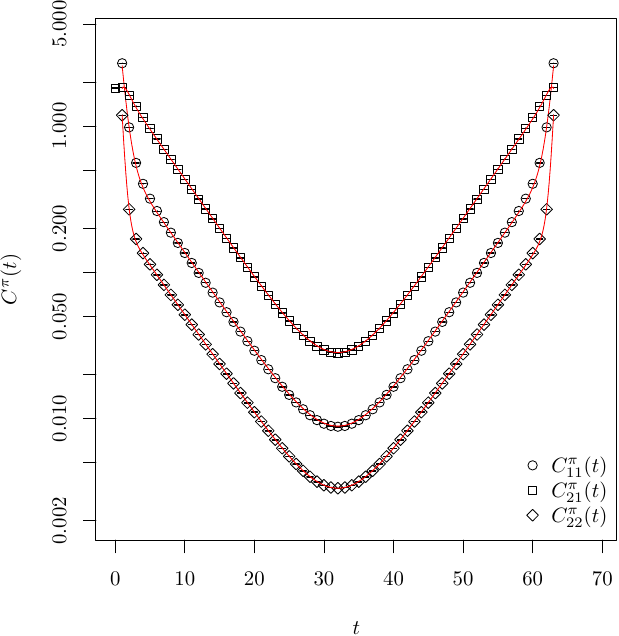}
	\caption{Pion correlator data together with the reconstruction from
		the THC method (red lines) using eq.~\eqref{eq:get_X_sym} and the
		truncation $k=6$. Error bands on the fitting curves are plotted
		but not visible at this scale. The elements
		$C^\pi_{21}=C^\pi_{12}$ and $C^\pi_{22}$ are shifted by a factor $3$
		and $1/3$, respectively, for better legibility.} 
	\label{fig:pion_correlator}
\end{figure}

\begin{figure}
	\centering
	\includegraphics[width=\linewidth,page=12]{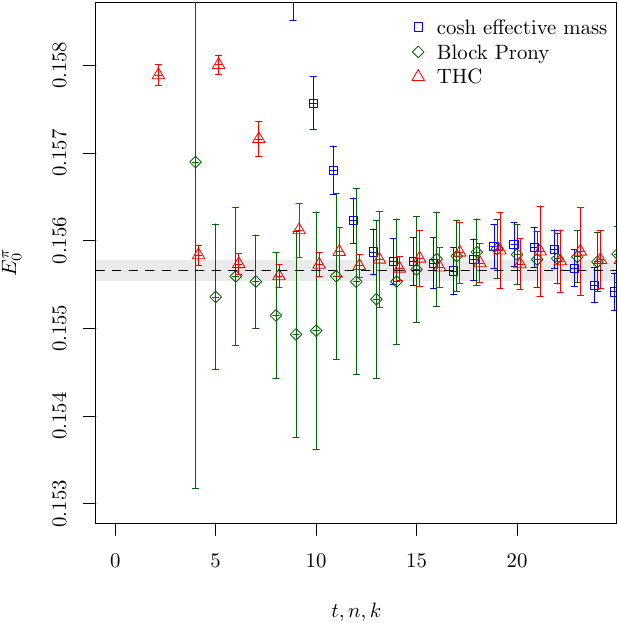}
	\caption{Ground state energy estimators (in lattice units) for the pion using the cosh-type effective mass on the $C^\pi_{11}(t)$ component of the correlator as well as block Prony and THC using the full $2\times 2$ correlator matrix. Convergence is shown as a function of $t$, $n$, or $k$, respectively. The dashed line with error band stems from Ref.~\cite{Ottnad:2012fv} and serves as a benchmark. Note that the Prony method required a double bootstrap analysis in order to get the errors under control while the effective mass and THC methods rely on simple bootstrap estimators.}
	\label{fig:pion_results}
\end{figure}

In figure~\ref{fig:pion_correlator} we show the reconstructed elements
of the correlator matrix (continuous red lines) based on the cut at
$k=6$ compared with the original data. We observe an overall excellent
description of the data. Finally, in figure~\ref{fig:pion_results} the
estimate of the ground state energy level from THC as a function of the
cut-off $k$ is compared to Block Prony as a function of the Hankel matrix
dimension $n$, as well as the $\cosh$ effective mass as a function of
$t$. The reference value with uncertainty is plotted as the horizontal
line with shaded error band. 

As expected from the discussion of figure~\ref{fig:pion_eigenvalues},
THC agrees with the reference value at $k=6$, where we obtain the
ground state energy as  $E^\pi = 0.15575(12)$. Since the correlator is
symmetric in Euclidean time, odd $k$-values converge later, and for
$k\geq9$ on all THC estimates agree with the reference value and its
uncertainty. The errors of the THC estimate increase again for larger
$k$-values since we start to mix noise modes back into the
reconstruction.

Block Prony on the other hand leads to significantly larger
uncertainties than THC, but also agrees with the reference value from
$n\geq5$ on. The $\cosh$ effective mass converges from $t=14$ on, with
slightly larger point errors than THC. But this uncertainty is not
comparable to THC, because much more information goes into the THC
estimate at $k=6$ than into the $\cosh$ effective mass at $t=14$.

\begin{figure}
	\centering
	\includegraphics[width=\linewidth,page=5]{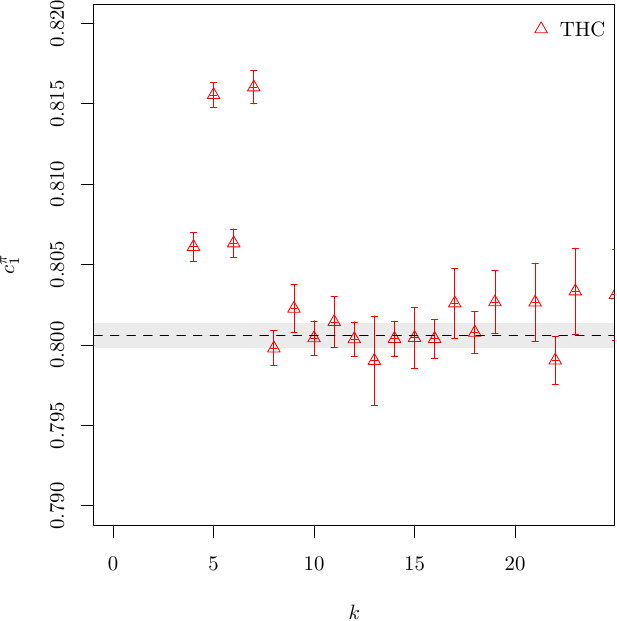}
	\caption{Ground state matrix element (vectorised coefficient
          $c_{1,1}$ from eq.~\eqref{eq:corr_matrix}) estimator for the
          pion the THC algorithm~\ref{alg:coefficients} on the full
          $2\times 2$ correlator matrix. Convergence is shown as a
          function of $k$. The dashed line with error band stems from
          a $\cosh$-fit (see text) and serves as a benchmark.}
	\label{fig:pion_coefficients}
\end{figure}

Since the pion correlators do not suffer from the StN problem, it is
possible to also investigate the convergence of the coefficients in
$k$. For $k=6$, for example, the coefficient is estimated as $c_1 =
0.8063(9)$, for $k=8$ we find $E^\pi = 0.15561(13)$ and $c_1 =
0.7998(11)$, see figure~\ref{fig:pion_coefficients}.
The coefficient $c_1$ extracted from a $\cosh$-fit to the correlator
element $C^\pi_{11}$ only from $t_1=13$ to $t_2=32$ reads $c_1=0.8006(8)$.
Thus, the coefficients extracted using THC appear to converge two or
so timeslices later than the energies. However, once converged
excellent agreement is found between the reference fit value and
THC. We remark that the coefficient $c_1$ is for this ensemble
directly related to the pion decay constant $f_\pi$ on this ensemble.

We remark that the default weights eq.~\eqref{eq:diag_weight_mat}
correspond for the pion to constant relative errors. Using these
should thus put equal weight to all correlator data points. We observe
empirically that for the pion $\Omega=\mathds{1}$ works equally well
for the ground state. This is understandable, since the ground state is
mainly determined at large times, which get relatively higher weighted
by $t$ independent weights compared to the default weights.

\subsubsection{$\omega$ Meson Correlator}

Next we investigate a correlator matrix where the signal-to-noise
ratio is exponentially decreasing in Eu\-cli\-de\-an time, which is actually
the usual situation in LQCD. For this we study a correlator matrix
$C^\omega$ of size $4\times4$ based on $201$ measurements taken from
Ref.~\cite{Yan:2024gwp}, which was used in a study of the $\omega$
meson from LQCD. 
It is based on an ensemble generated by the Chinese LQCD collaboration and is denoted
F48P30. For the generation $N_f=2+1$ Wilson clover fermions have been
used. The pion mass value of this ensemble corresponds to about
$300\ \mathrm{MeV}$. For more details on the ensembles we refer to
Ref.~\cite{CLQCD:2023sdb} and for how the correlator
matrix was estimated to Ref.~\cite{Yan:2024gwp}, the details
go beyond the scope of this paper. 
However, $C^\omega$ is again real, symmetric and symmetric in
time. And while $201$ measurements may sound little, an all-to-all
estimator was used which leads to reasonable statistical
uncertainties~\cite{Yan:2024gwp}. 

In Ref.~\cite{Yan:2024gwp}, $C^\omega$ was analysed using the original
GEVP method and the following four energy levels where found (in
lattice units):
\begin{equation}
  \label{eq:omegaEs}
  \begin{split}
    E^\omega_1 = 0.327(5)\,,\quad &E^\omega_2 = 0.512(4)\,,\\
    E^\omega_3 = 0.584(2)\,,\quad &E^\omega_4 = 0.615(3)\,.\\
  \end{split}
\end{equation}
From these four, only the lowest two are used in
Ref.~\cite{Yan:2024gwp} for the analysis. Since the matrix is only
$4\times4$, it is also expected that the fourth eigenvalue has large
systematic uncertainty, which might affect $E^\omega_3$ as well.

\begin{figure}
	\centering
	\includegraphics[width=\linewidth,page=1]{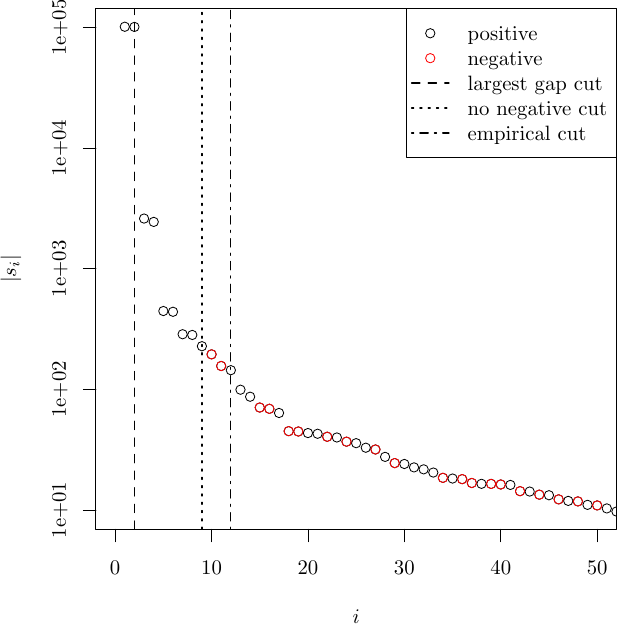}
	\caption{Modulus of the eigenvalues $s_i$ of the full Hankel
          matrix $\tilde H$ defined by
          eq.~\eqref{eq:weighted_full_Hankel} built from the $\omega$
          correlator starting at $t_0=2$ and using the default
          weights~\eqref{eq:diag_weight_mat}. Negative eigenvalues are
          marked red. Natural candidates for the truncation $k$ are
          $k=2$ at the largest ratio
          $\left|\frac{s_{k}}{s_{k+1}}\right|$ (i.e.\ the largest gap
          on the log-scale) and $k=9$ right before the first negative
          eigenvalue $s_{k+1}\le 0$. Empirically, we find that $k=12$
          is better for the stable resolution of high excited
          states.}
	\label{fig:omega_eigenvalues}
\end{figure}

In figure~\ref{fig:omega_eigenvalues} we show the modulus of the
spectrum of the full Hankel matrix $\tilde H$ using default weights
eq~\eqref{eq:diag_weight_mat}. The interpretation of this figure is
less clear cut than the one of the pion, but the default cuts would be
at $k=2$ and $k=9$ for the logarithmic gap and the first negative
eigenvalue, respectively. While the latter of these cuts (and any
other $k\ge 4$) works well for estimating the ground state energy,
$k=12$ turns out to be the better choice for estimating also excited
energy levels. 

\begin{figure*}
	\centering
	\includegraphics[width=.45\linewidth,page=3]{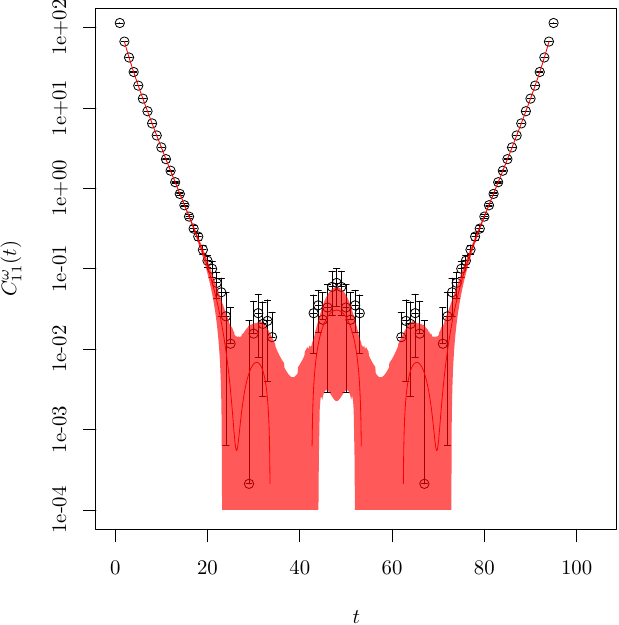}
	\hfill
	\includegraphics[width=.45\linewidth,page=5]{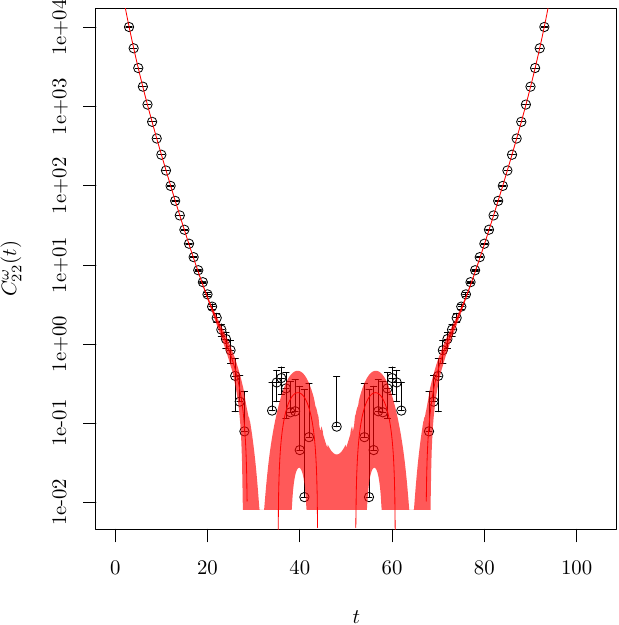}
	\caption{$\omega$ correlator data together with the reconstruction from the THC method (red lines) using eq.~\eqref{eq:get_X_sym} and the truncation $k=12$.}
	\label{fig:omega_correlator}
\end{figure*}

Using $k=12$ one obtains for instance the reconstructed correlator
elements $C_{11}$ and $C_{22}$ as shown in
figure~\ref{fig:omega_correlator}, where the agreement
with the original data is again excellent.
In this case we used the full coefficient matrices $\tilde c_{\alpha\beta,l}$ instead of the corresponding vector products $\tilde c_{\alpha\beta,l} \approx \tilde c_{\alpha,l} {\tilde c_{\beta,l}}^{\star}$. 
While for physical states the decomposition into the vectors $\tilde
c_{\alpha,l}$ was stable, it did not work well for the noise states with imaginary energies.
This does not come as a surprise since noisy states have no intrinsic structure and are not guaranteed to be rank-1 like the physical states in equation~\eqref{eq:corr_matrix}.
For involved analyses like this one, we recommend to verify the results plotting the matrix version as we did here. The coefficients $\tilde c_{\alpha,l}$ should then be extracted exclusively for the states of physical interest.

\begin{figure}
	\centering
	\includegraphics[width=\linewidth,page=2]{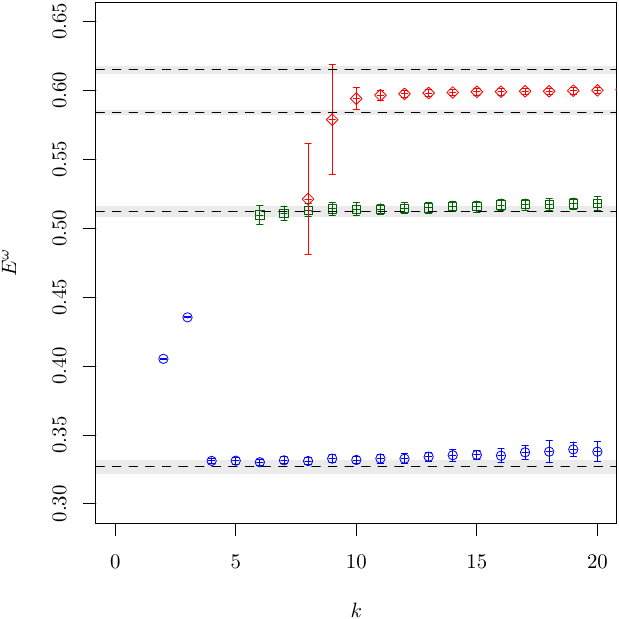}
	\caption{Energy estimators for the $\omega$ using the THC method on the full $4\times 4$ correlator matrix. Convergence is shown as a function of the truncation $k$. The dashed lines with error band stem from an independent analysis of the same data using a standard GEVM with plateau fits and serve as benchmarks. The THC errors rely on simple bootstrap estimators.}
	\label{fig:omega_results}
\end{figure}

In figure~\ref{fig:omega_results} we show the THC energy estimates for the
three lowest states as a function of $k$. The horizontal dashed lines
indicate the four energy values from eq.~\ref{eq:omegaEs}. For the
lowest two energy levels, we find
excellent agreement between reference and the THC estimates for large
enough $k$-values. In particular, from $k\geq12$ THC stably resolves
three states. However, THC does not find the equivalent of the fourths 
state found by the GEVP, but only one around $E=0.6$ (we recall that
third and fourth state were not used in Ref.~\cite{Yan:2024gwp}.)

As mentioned above, since the GEVP is based on a $4\times4$ matrix,
maximally four states can be resolved, and the higher states receive
most of the corrections due to unresolved states in the tower of
exponentials. Thus, it seems likely that the state estimated by the
THC is less affected by systematics. Here, a GEVP analysis with a
larger operator basis would be highly interesting.

Let us also point out that while in the GEVP so-called thermal
pollution states need to be carefully removed during analysis, the THC
does not require this. Thermal pollution states will be resolved as
separate states by the THC, if statistical accuracy suffices.

\subsubsection{Nucleon Correlator}

Finally, we also test the THC method on a single nucleon
correlator, a state badly affected by the StN problem.
$C^N$ was computed on $401$ configuration of a
$N_f = 2+1+1$ Wilson twisted mass
lattice ensemble of size $80^3\times160$ denoted cC211.06.80 in
Ref.~\cite{ExtendedTwistedMass:2021gbo}.
The pion mass of this ensemble corresponds to about
$134\ \mathrm{MeV}$, at a lattice spacing of
$a = 0.06860(20)\ \mathrm{fm}$~\cite{ExtendedTwistedMass:2021gbo}.

Details for the construction of the single Euclidean correlation
function $C^N(t)$ can be found 
in Ref.~\cite{ExtendedTwistedMass:2021gbo}, from which we also take
the reference value $E^N = 0.3261(11)$ (in lattice units) from a two state fit with
statistical error only.
$C^N$ is real, but not symmetric in time; the back-propagating
state represents the parity partner of the nucleon.\footnote{We thank
S.~Bacchio for communicating the first and second excited state from a
three state fit to us.}

\begin{figure}
  \centering
  \includegraphics[width=\linewidth, page=2]{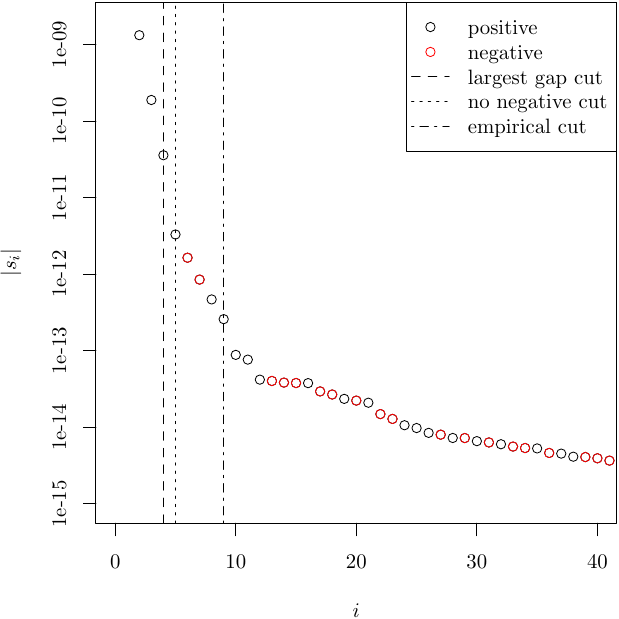}
  \caption{Modulus of the eigenvalues $s_i$ of the full Hankel
    matrix $\tilde H$ defined by
    eq.~\eqref{eq:weighted_full_Hankel} built from the nucleon
    correlator starting at $t_0=1$ and using constant weights
    $\Omega=\mathds{1}$. For better visibility we show
    only eigenvalues up to index $i=41$.
    Negative eigenvalues are
    marked red. Here, the default choice for the cuts would be
    $k=4$ and $k=5$, but the relevant cut is at $k=9$, see text.}
  \label{fig:proton_eigenvalues}
\end{figure}

\begin{figure*}
  \centering
  \includegraphics[width=.45\linewidth,page=4]{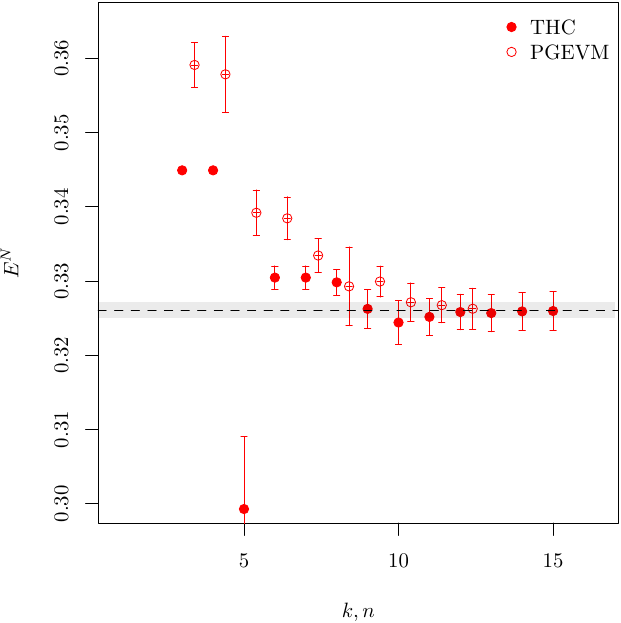}
  \hfill
  \includegraphics[width=.45\linewidth,page=5]{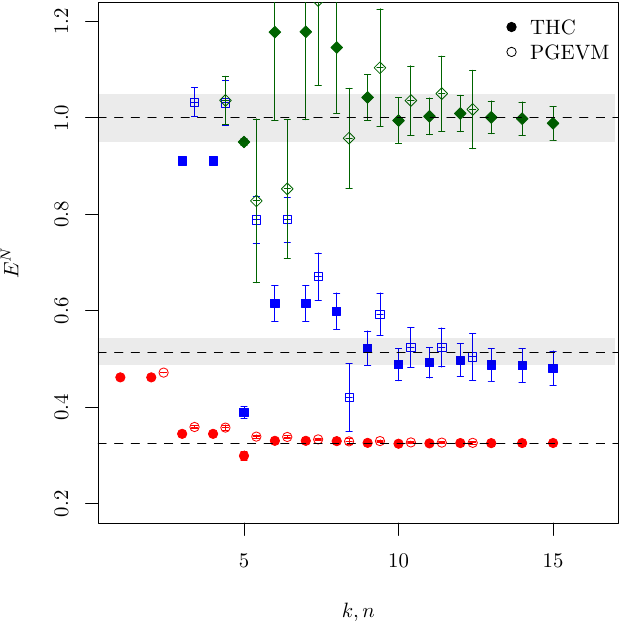}
  \caption{Energy estimates from the nucleon correlator $C^N$
    comparing THC with $t_0=1$, PGEVM with $t_0=1$ and a three state
    fit (horizontal lines). In the left panel only the ground state is
  compared for better legibility. In the right panel we compare the
  first three states.}
  \label{fig:proton_results}
\end{figure*}

In figure~\ref{fig:proton_eigenvalues} we show the eigenvalues of the full
Hankel matrix $\tilde H$ defined by
eq.~\eqref{eq:weighted_full_Hankel} for the nucleon correlator with
$t_0=1$ with constant weights $\Omega=\mathds{1}$.
We have verified that including the multiplicity factors $\#(t)$ from eq.~\eqref{eq:multiplicity} but not the errors $\sigma$ leads to practically identical results.
Our default cuts
would suggest to retain only $k=4$ or $k=5$ 
eigenvalues. However, it turns out that the two negative eigenvalues
with the largest modulus are important to describe the correlator: the
reason is that $0< C^N(0) < C^N(1)$, which requires a negative mode in
$\tilde H$. This negative mode has impact beyond $t=0$ (which is
typically unphysical due to contact terms) since the interpolating
operators are smeared~\cite{ExtendedTwistedMass:2021gbo}.
Thus, a good choice is $k=9$ indicated by the logarithmic gap between
the ninth and tenth eigenvalues.

In the left panel of figure~\ref{fig:proton_results} we compare the
extraction of the ground state energy level $E^N_1$ between THC and
PGEVM. The dashed line represents the reference value from
Ref.~\cite{ExtendedTwistedMass:2021gbo}. 
Confirming the discussion from the last paragraph, the THC result
converges only for $k\geq9$, but shows excellent agreement to the
two-state fit reference result. We recall that the reference value has
a statistical only uncertainty.
The relevance of the negative eigenvalues is further indicated by the
THC estimate of $E^N$ at $k=5$, which is significantly lower than the
surrounding values: the first $5$ eigenvalues are not sufficient to
describe the correlator accurately.
For the PGEVM, we had to remove generalised eigenvalues larger than
$0.9$ by hand (which amounts to noise removal by hand) to obtain the
results displayed in the figure. But with this to some extend
arbitrary cut we obtain agreement to THC albeit slightly larger
uncertainties.

In the right panel of this figure we again compare THC with PGEVM and
reference values, but for the three lowest states in the
correlator. The dashed lines are the reference estimates from a three
state fit to the correlator. Comparing THC with PGEVM, THC appears
more stable, in particular for $k\geq9$, from which on all three states
appear converged. The PGEVM results agree with the THC results in
principle, but appear less stable and show again slightly larger
uncertainties. THC also agrees with the reference values for the first
and second excited states with comparable or slightly smaller
uncertainties. Again, the three state fit uncertainties are
statistical only and expected to increase once
systematic effects are accounted for. 

We mentioned already at the beginning of this subsection that for the
nucleon data we use double bootstrap (for both THC and PGEVM). The
reason for this choice is that the uncertainties of the two excited
states turn out to be unrealistic otherwise. This means in practice a factor
three to five times larger error on $E^N_{2,3}$, while the ground
state is not affected. This might be related to the fact that a single
correlator is investigated for the nucleon. A corresponding analysis
with a correlator matrix would be very interesting, but goes beyond
the scope of this paper.

However, this topic might be related to the fact that for the nucleon
we had to use constant weights $\Omega=\mathds{1}$
as mentioned before. Using default weights as defined in
eq.~\eqref{eq:diag_weight_mat} with reasonably small truncations $k\le20$
leads to a bias in the results for
the nucleon. We attribute this to the exponentially deteriorating StN
ratio, but currently lack a detailed understanding for this observation.

\begin{figure}
  \centering
  \includegraphics[width=\linewidth, page=3]{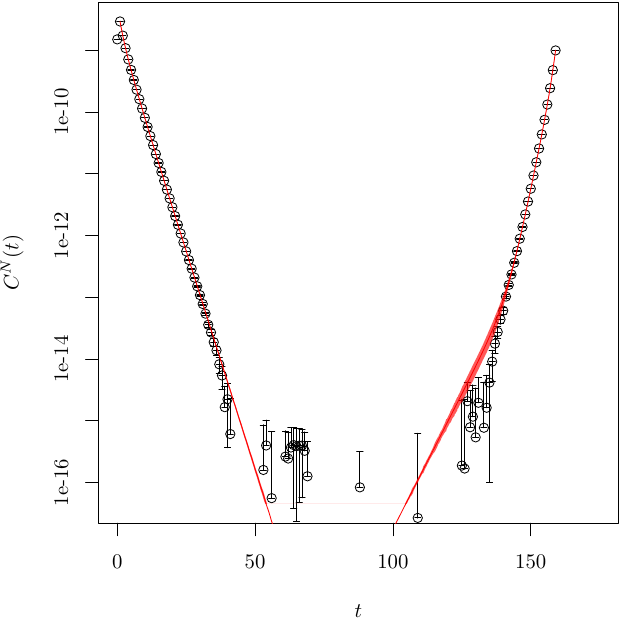}
  \caption{Reconstructed nucleon correlator for $k=11$ as red
    continuous line confronted with the data for $C^N(t)$.}
  \label{fig:proton_correlator}
\end{figure}

Finally, we show the reconstructed nucleon correlator for truncation
$k=11$ in figure~\ref{fig:proton_correlator} as a continuous line versus
the data for $C^N(t)$. In the region $40\leq t\leq140$ the data is
compatible with zero. The reconstructed correlator nicely describes
the data. Also the difference in energy between the forward and
backward propagating state is clearly visible.

\section{Conclusion}

In this paper we have introduced a novel algebraic method called
Truncated Hankel Correlator (THC) method for the spectral analysis of
Euclidean correlation functions.
Algorithm~\ref{alg:basic_THC} summarises the THC method in its simplest form.
We have argued that this method
produces an optimal low-rank approximation of the Hankel matrix
eq.~(\ref{eq:maximal_rect_hankel}) which is then used to determine
coefficients and energy levels from noisy estimates of a matrix-valued
correlator eq.~\eqref{eq:corr_matrix}.

Weights, informed by the uncertainties of individual measurements, can be included into the approximation (alg.~\ref{alg:full_THC}).
At the same time, the optimisation of the exact corresponding $\chi^2$-function~\eqref{eq:chi_sq} is proven to be exponentially hard in the number of states.
This puts the THC method right at the sweet spot of great accuracy with polynomial runtime.
We have also shown that commonly used alternative algebraic methods
like the GEVP, the Prony GEVP, Lanczos, and their block versions are
sub-optimal special cases of THC.

The THC method has only a single parameter $k$, which represents the
number of states retained in the approximated Hankel matrix. When
chosen large enough to cover all the physically relevant states in the
correlation function, results become independent of $k$ and indicate
convergence. Once $k$ has been appropriately chosen, no additional
fits or averaging procedures are required. In particular there is no
additional systematic effect expected due to the choice of $k$ once
convergence is observed.
Overall, only minimal human oversight and time are required to obtain reliable and accurate results with the THC method.

The parameter $k$ can be tuned based on the spectrum of the original
Hankel matrix. We have proposed different criteria derived from this
spectrum: cut after the largest logarithmic gap, or when the first
negative eigenvalue appears. However, for some of the real world data
examples it turned out that the performance of the THC method was improved
by a cut at larger $k$, which can always be estimated from the
convergence of the states.

We have tested the THC method on synthetic data first and found that
it converges faster and more reliably than e.g.\ Lanczos or
PGEVM. Moreover, time-symmetric correlators can be treated such that
this symmetry is exactly implemented in the observed spectrum, which
is not possible for PGEVM (and Lanczos). Next we tested THC for
lattice quantum chromodynamics data, more specifically for the case of
the pion, the $\omega$ meson, and the nucleon. While for the pion and
$\omega$ the correlation functions were matrix valued, for the nucleon
we considered a scalar correlator. For all three hadrons we could
reproduce reference values from other methods or the literature with
excellent agreement. For the case of the pion we also showed that the
coefficients or amplitudes of the single exponential terms can be
estimated with competitive uncertainties, even if convergence is
observed for slightly larger $k$-values than needed for the energy
levels.

Both, the $\omega$ meson and the nucleon are affected by the signal-to-noise (StN) problem.
We emphasise that the StN problem has not been solved.
However, we find that the THC method is very robust against noise and thermal pollutions.
That is, no non-trivial filtering is required to identify physical states on top of the truncation and no contamination or bias from negative energies if observed.

In short, we highly encourage the wide-spread use of THC.

In the future it will be interesting to better understand the spectrum
of random Hankel matrices in order to possibly give a theoretically
sounder cutting criterion. Moreover, the relation between matrix
dimension of the correlator and the stability in the error estimates
for excited states will be interesting to understand better. 

~\\%
\section*{Acknowledgments}
We thank M.\ Garofalo, C.\ Groß, E.\ Gull, T.\ Hartung, B.\ Kostrzewa, T.\ Luu, S.\ Mallach, G.\ Pederiva, M.\ Rodekamp, and A.\ Sen for insightful discussions.
We thank M.\ Garofalo, M.\ Mai, and H.\ Yan for providing $\omega$
meson data and S.\ Bacchio and C.\ Alexandrou for proton data.
This work was funded by the Deutsche Forschungsgemeinschaft (DFG, German Research Foundation) as part of the
CRC 1639 NuMeriQS – Project number 511713970.
The open source software packages R~\cite{R:2019} and hadron~\cite{hadron:2024},
now including a full implementation of the THC algorithm,
have been used.

\appendix
\section{Proof of convergence for block PGEVM}\label{sec:block_Prony_proof}

For completeness sake, we provide a proof that the generalisation of Prony's method using blocks of correlator matrices works in the expected way. The proof proceeds analogously to the scalar version in Ref.~\cite{Fischer:2020bgv}.

Consider a $d\times d$ correlator matrix $C_{\alpha\beta}(t)$ as in equation~\eqref{eq:corr_matrix}.
Now define the matrix $H$ at $t$ by
\begin{equation}
  \begin{split}
    H_{ij,\alpha\beta}(t) &= C_{\alpha\beta}(t+i\Delta+j\Delta)\\
    &=\sum_{l=0}^{n-1}e^{-E_lt}\ e^{-E_l i \Delta}\ e^{-E_l j \Delta}
    c_{\alpha,l}c_{\beta,l}^\star\\
    &=\sum_{l=0}^{n-1}e^{-E_lt}\xi_{l}^{i\alpha}\xi_{l}^{j\beta\star}\,,
  \end{split}
\end{equation}
i.e.\ a Hankel matrix of correlator matrices, 
with total dimension $n=N d$ and indices running like
\[
\begin{split}
  l &= 0, \ldots, n-1\,,\\
  i,j &= 0, \ldots, N-1\,,\\
  \alpha,\beta &= 0, \ldots, d-1\,.
\end{split}
\]
In addition we have defined $\xi$ (a generalisation of the Vandermonde matrix $\chi$) as follows
\begin{equation}
  \xi_{l}^{i\alpha} = e^{-E_l i \Delta}\, c_{\alpha,l}\,.
\end{equation}
Now, with $l$ indexing the columns and $id+\alpha$ the columns, $\xi$
can be considered a square matrix of the form 
\begin{equation}
	\label{eq:chimatrix}
	\xi =
	\begin{pmatrix}
		c_{0,0} & \cdots & c_{0,n-1} \\
		\vdots & \ddots & \vdots\\
		c_{d-1,0} & \cdots & c_{d-1,n-1}\\
		c_{0,0}e^{-E_0\Delta} & \cdots & c_{0,n-1}e^{-E_{n-1}\Delta} \\
		\vdots & \ddots & \vdots \\
		c_{d-1,0}e^{-E_0\Delta} & \cdots & c_{d-1,n-1}e^{-E_{n-1}\Delta}\\
		\vdots & \ddots & \vdots \\
		c_{d-1,0}e^{-(N-1)E_0\Delta} & \cdots & c_{d-1,n-1}e^{-(N-1)E_{n-1}\Delta}\\
	\end{pmatrix}\,.
\end{equation}
Under reasonable assumptions for the coefficients $c_{\alpha,l}$, this
matrix is invertible. Thus, we can define dual vectors $u_k$ with 
\begin{equation}
  \label{eq:orthogonality}
  (u_k, \xi_l) = \sum_{i,\alpha}
  (u_k^\star)^{(i,\alpha)}(\xi_l)^{(i,\alpha)} = \delta_{kl}
\end{equation}
and find
\begin{equation}
  \label{eq:eigenvectors}
  \begin{split}
    H(t) u_l &= \sum_{k=0}^{n-1} e^{-E_k t}\xi_k\xi_k^\dagger u_l\\
    &= e^{-E_l t} \xi_l = e^{-E_l(t-t_0)}e^{-E_l t_0}\xi_l\\
    &=e^{-E_l(t-t_0)} H(t_0)u_l\,.
  \end{split}
\end{equation}
Therefore,
\begin{equation}
  \Lambda_l(t,t_0) = e^{-E_l(t-t_0)}
\end{equation}
is an eigenvalue of the generalised eigenvalue problem
\begin{equation}
  H(t)\, v_l = \Lambda_l\ H(t_0)\, v_l
\end{equation}
and $v_l\propto u_l$ for all $l$. From Eq.~\ref{eq:eigenvectors} one
can observe 
\begin{equation}
  \xi_l = e^{E_l t} H(t) u_l\,,
\end{equation}
one of the columns of the $\xi$ matrix Eq.~\ref{eq:chimatrix}, and with
Eq.~\ref{eq:orthogonality} 
\begin{equation}
  (u_l, H(t) u_k) = e^{-E_l t} \delta_{lk}\,.
\end{equation}
Now, assuming $v_l$ have been determined by solving the GEVP, there is
an additional proportionality factor to take care of. Set
\begin{equation}
  a_l = \sqrt{(v_l, H(t) v_l)}\,,
\end{equation}
then with $u_l = v_l\,\exp(-E_l t/2)/a_l$:
\begin{equation}
  (u_k, H(t) u_l) = \left( \frac{v_k}{e^{E_k t/2}a_k}, H(t)
  \frac{v_l}{e^{E_l t/2}a_l} \right) = e^{-E_l t} \delta_{lk}\,.
\end{equation}
Thus, one obtains the matrix elements $c_{\alpha,l}=\langle
0|O_\alpha|l\rangle$ 
\begin{equation}
  \begin{split}
    c_{\alpha,l} &= e^{E_l i \Delta}\xi_{l}^{i\alpha} =
    e^{E_l i \Delta} e^{E_lt} (H(t) u_l)^{(i,\alpha)}\\
    &=\frac{1}{a_l} e^{E_l (t/2 + i \Delta)}  (H(t) v_l)^{(i,\alpha)}\,,
  \end{split}\label{eq:calculate_c_ka}
\end{equation}
for all $i=0,\ldots, N-1$, since the $c_{\alpha,l}$ are
independent of $t$, $i$, and $\Delta$. 

\printbibliography

@article{ETM:2008zte,
    author = "Boucaud, Philippe and others",
    collaboration = "ETM",
    title = "{Dynamical Twisted Mass Fermions with Light Quarks: Simulation and Analysis Details}",
    eprint = "0803.0224",
    archivePrefix = "arXiv",
    primaryClass = "hep-lat",
    reportNumber = "DESY-08-022, IFIC-08-08, FTUV-08-1302, MS-TP-08-3, RM3-TH-08-5, ROM2F-2008-04, SFB-CPP-08-17, LTH-762",
    doi = "10.1016/j.cpc.2008.06.013",
    journal = "Comput. Phys. Commun.",
    volume = "179",
    pages = "695--715",
    year = "2008"
}

@article{Lacock:1994qx,
    author = "Lacock, P. and McKerrell, A. and Michael, Christopher and Stopher, I. M. and Stephenson, P. W.",
    collaboration = "UKQCD",
    title = "{Efficient hadronic operators in lattice gauge theory}",
    eprint = "hep-lat/9412079",
    archivePrefix = "arXiv",
    reportNumber = "LTH-344, SWAT-58",
    doi = "10.1103/PhysRevD.51.6403",
    journal = "Phys. Rev. D",
    volume = "51",
    pages = "6403--6410",
    year = "1995"
}

@article{Ostmeyer:2024qgu,
    author = "Ostmeyer, Johann and Sen, Aniket and Urbach, Carsten",
    title = "{On the equivalence of Prony and Lanczos methods for Euclidean correlation functions}",
    eprint = "2411.14981",
    archivePrefix = "arXiv",
    primaryClass = "hep-lat",
    month = "11",
    year = "2024"
}

@article{Ottnad:2012fv,
    author = "Ottnad, Konstantin and Michael, C. and Reker, S. and Urbach, C. and Michael, Chris and Reker, Siebren and Urbach, Carsten",
    collaboration = "ETM",
    title = "{{\textbackslash}eta and {\textbackslash}eta' mesons from Nf=2+1+1 twisted mass lattice QCD}",
    eprint = "1206.6719",
    archivePrefix = "arXiv",
    primaryClass = "hep-lat",
    doi = "10.1007/JHEP11(2012)048",
    journal = "JHEP",
    volume = "11",
    pages = "048",
    year = "2012"
}

@article{Baron:2010bv,
    author = "Baron, R. and others",
    title = "{Light hadrons from lattice QCD with light (u,d), strange and charm dynamical quarks}",
    eprint = "1004.5284",
    archivePrefix = "arXiv",
    primaryClass = "hep-lat",
    reportNumber = "DESY-10-054, HU-EP-10-18, IFIC-10-11, SFB-CPP-10-29, LPT-ORSAY-10-28, LTH873, LPSC1042, MS-TP-10-09, ROM2F-2010-08",
    doi = "10.1007/JHEP06(2010)111",
    journal = "JHEP",
    volume = "06",
    pages = "111",
    year = "2010"
}

@article{Yan:2024gwp,
    author = "Yan, Haobo and Mai, Maxim and Garofalo, Marco and Mei{\ss}ner, Ulf-G. and Liu, Chuan and Liu, Liuming and Urbach, Carsten",
    title = "{{\ensuremath{\omega}} Meson from Lattice QCD}",
    eprint = "2407.16659",
    archivePrefix = "arXiv",
    primaryClass = "hep-lat",
    doi = "10.1103/PhysRevLett.133.211906",
    journal = "Phys. Rev. Lett.",
    volume = "133",
    number = "21",
    pages = "211906",
    year = "2024"
}

@article{CLQCD:2023sdb,
    author = "Hu, Zhi-Cheng and others",
    collaboration = "CLQCD",
    title = "{Quark masses and low-energy constants in the continuum from the tadpole-improved clover ensembles}",
    eprint = "2310.00814",
    archivePrefix = "arXiv",
    primaryClass = "hep-lat",
    doi = "10.1103/PhysRevD.109.054507",
    journal = "Phys. Rev. D",
    volume = "109",
    number = "5",
    pages = "054507",
    year = "2024"
}

@article{ExtendedTwistedMass:2021gbo,
    author = "Alexandrou, C. and others",
    collaboration = "Extended Twisted Mass",
    title = "{Quark masses using twisted-mass fermion gauge ensembles}",
    eprint = "2104.13408",
    archivePrefix = "arXiv",
    primaryClass = "hep-lat",
    doi = "10.1103/PhysRevD.104.074515",
    journal = "Phys. Rev. D",
    volume = "104",
    number = "7",
    pages = "074515",
    year = "2021"
}

@article{Ottnad:2017bjt,
    author = "Ottnad, Konstantin and Urbach, Carsten",
    collaboration = "ETM",
    title = "{Flavor-singlet meson decay constants from $N_f=2+1+1$ twisted mass lattice QCD}",
    eprint = "1710.07986",
    archivePrefix = "arXiv",
    primaryClass = "hep-lat",
    doi = "10.1103/PhysRevD.97.054508",
    journal = "Phys. Rev. D",
    volume = "97",
    number = "5",
    pages = "054508",
    year = "2018"
}

@article{Abbott:2025yhm,
    author = "Abbott, Ryan and Hackett, Daniel C. and Fleming, George T. and Pefkou, Dimitra A. and Wagman, Michael L.",
    title = "{Filtered Rayleigh-Ritz is all you need}",
    eprint = "2503.17357",
    archivePrefix = "arXiv",
    primaryClass = "hep-lat",
    reportNumber = "FERMILAB-PUB-25-0131-T, MIT-CTP/5849",
    month = "3",
    year = "2025",
    url={https://arxiv.org/abs/2503.17357}, 
}

@Manual{R:2019,
    title = {R: A Language and Environment for Statistical Computing},
    author = {{R Core Team}},
    organization = {R Foundation for Statistical Computing},
    address = {Vienna, Austria},
    year = {2019},
    url = {https://www.R-project.org/},
}

@article{hadron:2024,
    author       = {Kostrzewa, Bartosz and
    Ostmeyer, Johann and
    Ueding, Martin and
    Urbach, Carsten and
    Schlage, Nikolas and
    Werner, Markus and
    Pittler, Ferenc and
    Fischer, Matthias and
    Lubicz, Vittorio},
    title        = "{hadron: R Package for Statistical Methods to
    Extract (Hadronic) Quantities from Correlation
    Functions in Monte Carlo Simulations}",
    month        = oct,
    year         = 2025,
    publisher    = {Zenodo},
    doi          = {10.5281/zenodo.17375156},
    url          = {https://doi.org/10.5281/zenodo.17375156},
    addendum = "\url{https://github.com/HISKP-LQCD/hadron}",
    version = {3.4.1, {R} package}
}

@article{Wagman:2024rid,
    author = "Wagman, Michael L.",
    title = "{Lanczos, the transfer matrix, and the signal-to-noise problem}",
    eprint = "2406.20009",
    archivePrefix = "arXiv",
    primaryClass = "hep-lat",
    reportNumber = "FERMILAB-PUB-24-0320-T",
    month = "6",
    year = "2024"
}

@article{Fischer:2020bgv,
    author = "Fischer, Matthias and Kostrzewa, Bartosz and Ostmeyer, Johann and Ottnad, Konstantin and Ueding, Martin and Urbach, Carsten",
    title = "{On the generalised eigenvalue method and its relation to Prony and generalised pencil of function methods}",
    eprint = "2004.10472",
    archivePrefix = "arXiv",
    primaryClass = "hep-lat",
    doi = "10.1140/epja/s10050-020-00205-w",
    journal = "Eur. Phys. J. A",
    volume = "56",
    number = "8",
    pages = "206",
    year = "2020"
}

@InCollection{Lepage:1989,
  author =       {Lepage, G. P.},
  title =        {The Analysis Of Algorithms For Lattice Field Theory},
  booktitle =    {},
  key =       {QCD161:T45:1989},
  year =      {1989},
  note =      {Invited lectures given at TASI’89 Summer School, Boulder, CO, Jun 4-30, 1989. Published in Boulder ASI 1989:97-120 (QCD161:T45:1989)},
}

@article{Michael:1982gb,
      author         = "Michael, Christopher and Teasdale, I.",
      title          = "{Extracting Glueball Masses From Lattice {QCD}}",
      journal        = "Nucl. Phys.",
      volume         = "B215",
      year           = "1983",
      pages          = "433-446",
      doi            = "10.1016/0550-3213(83)90674-0",
      reportNumber   = "LTH 96",
      SLACcitation   = "%%CITATION = NUPHA,B215,433;%%"
}

@article{Luscher:1990ck,
      author         = "Lüscher, Martin and Wolff, Ulli",
      title          = "{How to Calculate the Elastic Scattering Matrix in
                        Two-dimensional Quantum Field Theories by Numerical
                        Simulation}",
      journal        = "Nucl. Phys.",
      volume         = "B339",
      year           = "1990",
      pages          = "222-252",
      doi            = "10.1016/0550-3213(90)90540-T",
      reportNumber   = "DESY-90-010",
      SLACcitation   = "%%CITATION = NUPHA,B339,222;%%"
}

@article{Blossier:2009kd,
      author         = "Blossier, Benoit and Della Morte, Michele and von Hippel,
                        Georg and Mendes, Tereza and Sommer, Rainer",
      title          = "{On the generalized eigenvalue method for energies and
                        matrix elements in lattice field theory}",
      journal        = "JHEP",
      volume         = "04",
      year           = "2009",
      pages          = "094",
      doi            = "10.1088/1126-6708/2009/04/094",
      eprint         = "0902.1265",
      archivePrefix  = "arXiv",
      primaryClass   = "hep-lat",
      reportNumber   = "DESY-09-014, SFB-CPP-09-10, MKPH-T-09-01,
                        LPT-ORSAY-09-05",
      SLACcitation   = "%%CITATION = ARXIV:0902.1265;%%"
}

@article{Prony:1795,
    author   = "de Prony, G. R.",
    journal  = "Journal de l’cole Polytechnique",
    volume = "1",
    number = "22",
    pages = "24-76",
    year = "1795"
}

@article{Cushman:2019hfh,
      author         = "Cushman, Kimmy K. and Fleming, George T.",
      title          = "{Automated label flows for excited states of correlation
                        functions in lattice gauge theory}",
      year           = "2019",
      eprint         = "1912.08205",
      archivePrefix  = "arXiv",
      primaryClass   = "hep-lat",
      SLACcitation   = "%%CITATION = ARXIV:1912.08205;%%"
}

@article{Cushman:2019tcv,
      author         = "Cushman, Kimmy K. and Fleming, George T.",
      title          = "{Prony methods for extracting excited states}",
      booktitle      = "{Proceedings, 36th International Symposium on Lattice
                        Field Theory (Lattice 2018): East Lansing, MI, United
                        States, July 22-28, 2018}",
      journal        = "PoS",
      volume         = "LATTICE2018",
      year           = "2019",
      pages          = "297",
      doi            = "10.22323/1.334.0297",
      eprint         = "1902.10695",
      archivePrefix  = "arXiv",
      primaryClass   = "hep-lat",
      SLACcitation   = "%%CITATION = ARXIV:1902.10695;%%"
}

@inproceedings{Fleming:2004hs,
      author         = "Fleming, George Tamminga",
      title          = "{What can lattice QCD theorists learn from NMR
                        spectroscopists?}",
      booktitle      = "{QCD and numerical analysis III. Proceedings, 3rd
                        International Workshop, Edinburgh, UK, June 30-July 4,
                        2003}",
      url            = "http://www1.jlab.org/Ul/publications/view_pub.cfm?pub_id=5245",
      year           = "2004",
      pages          = "143-152",
      eprint         = "hep-lat/0403023",
      archivePrefix  = "arXiv",
      primaryClass   = "hep-lat",
      reportNumber   = "JLAB-THY-04-13",
      SLACcitation   = "%%CITATION = HEP-LAT/0403023;%%"
}

@inproceedings{Fleming:2023zml,
    author = "Fleming, George T.",
    title = "{Beyond Generalized Eigenvalues in Lattice Quantum Field Theory}",
    booktitle = "{40th International Symposium on Lattice Field Theory}",
    eprint = "2309.05111",
    archivePrefix = "arXiv",
    primaryClass = "hep-lat",
    reportNumber = "FERMILAB-PUB-23-495-T",
    month = "9",
    year = "2023"
}

@article{Ottnad:2017mzd,
    author = "Ottnad, Konstantin and Harris, Tim and Meyer, Harvey and von Hippel, Georg and Wilhelm, Jonas and Wittig, Hartmut",
    editor = "Della Morte, M. and Fritzsch, P. and G\'amiz S\'anchez, E. and Pena Ruano, C.",
    title = "{Nucleon average quark momentum fraction with $N_\mathrm{f}=2+1$ Wilson fermions}",
    eprint = "1710.07816",
    archivePrefix = "arXiv",
    primaryClass = "hep-lat",
    reportNumber = "MITP-17-066",
    doi = "10.1051/epjconf/201817506026",
    journal = "EPJ Web Conf.",
    volume = "175",
    pages = "06026",
    year = "2018"
}

@article{Romiti:2019qim,
    author = "Romiti, S. and Simula, S.",
    title = "{Extraction of multiple exponential signals from lattice correlation functions}",
    eprint = "1907.09926",
    archivePrefix = "arXiv",
    primaryClass = "hep-lat",
    doi = "10.1103/PhysRevD.100.054515",
    journal = "Phys. Rev. D",
    volume = "100",
    number = "5",
    pages = "054515",
    year = "2019"
}

@article{Beane:2009kya,
    author = "Beane, Silas R. and Detmold, William and Luu, Thomas C. and Orginos, Kostas and Parreno, Assumpta and Savage, Martin J. and Torok, Aaron and Walker-Loud, Andre",
    title = "{High Statistics Analysis using Anisotropic Clover Lattices: (I) Single Hadron Correlation Functions}",
    eprint = "0903.2990",
    archivePrefix = "arXiv",
    primaryClass = "hep-lat",
    reportNumber = "UNH-09-01, JLAB-THY-09-960, NT@UW-09-08, ICCUB-09-18, ATHENA-PUB-09-012",
    doi = "10.1103/PhysRevD.79.114502",
    journal = "Phys. Rev. D",
    volume = "79",
    pages = "114502",
    year = "2009"
}

@article{Zhang:2024mpr,
	title = "{Minimal pole representation and controlled analytic continuation of Matsubara response functions}",
	author = {Zhang, Lei and Gull, Emanuel},
	journal = {Phys. Rev. B},
	volume = {110},
	issue = {3},
	pages = {035154},
	numpages = {7},
	year = {2024},
	month = {Jul},
	publisher = {American Physical Society},
	doi = {10.1103/PhysRevB.110.035154},
	url = {https://link.aps.org/doi/10.1103/PhysRevB.110.035154}
}

@article{Beylkin:2005oao,
	author = {Beylkin, Gregory and Monzón, Lucas},
	year = {2005},
	month = {07},
	pages = {17-48},
	title = {On approximation of functions by exponential sums},
	volume = {19},
	journal = {Applied and Computational Harmonic Analysis},
	doi = {10.1016/j.acha.2005.01.003}
}

@article{Beylkin:2010abe,
	title = {Approximation by exponential sums revisited},
	journal = {Applied and Computational Harmonic Analysis},
	volume = {28},
	number = {2},
	pages = {131-149},
	year = {2010},
	note = {Special Issue on Continuous Wavelet Transform in Memory of Jean Morlet, Part I},
	issn = {1063-5203},
	doi = {https://doi.org/10.1016/j.acha.2009.08.011},
	url = {https://www.sciencedirect.com/science/article/pii/S1063520309000906},
	author = {Gregory Beylkin and Lucas Monzón},
}

@inproceedings{Srebro:2003wlra,
	title={Weighted Low-Rank Approximations},
	author={Srebro, Nathan and Jaakkola, Tommi},
	booktitle={Proceedings of the 20th International Conference on Machine Learning (ICML)},
	pages={720--727},
	year={2003},
	url_Paper={https://www.aaai.org/Library/ICML/2003/icml03-094.php}
}

@article{Gillis:2011lrma,
	author = {Gillis, Nicolas and Glineur, Fran\c{c}ois},
	title = "{Low-Rank Matrix Approximation with Weights or Missing Data Is NP-Hard}",
	journal = {SIAM Journal on Matrix Analysis and Applications},
	volume = {32},
	number = {4},
	pages = {1149-1165},
	year = {2011},
	doi = {10.1137/110820361},
	eprint = {https://doi.org/10.1137/110820361}
}

@article{Yu:2024ncm,
	author = "Yu, Yang and Kemper, Alexander F. and Yang, Chao and Gull, Emanuel",
	title = "{Denoising of imaginary time response functions with Hankel projections}",
	eprint = "2403.12349",
	archivePrefix = "arXiv",
	primaryClass = "cond-mat.str-el",
	doi = "10.1103/PhysRevResearch.6.L032042",
	journal = "Phys. Rev. Res.",
	volume = "6",
	number = "3",
	pages = "L032042",
	year = "2024"
}

@article{Chakraborty:2024exj,
    author = "Chakraborty, Debsubhra and Sood, Dhruv and Radhakrishnan, Archana and Mathur, Nilmani",
    title = "{Estimating energy levels from lattice QCD correlation functions using a transfer matrix formalism}",
    eprint = "2412.01900",
    archivePrefix = "arXiv",
    primaryClass = "hep-lat",
    reportNumber = "TIFR/TH/24-24",
    month = "12",
    year = "2024"
}

@article{ottaviani2014exact,
	title={Exact solutions in structured low-rank approximation},
	author={Ottaviani, Giorgio and Spaenlehauer, Pierre-Jean and Sturmfels, Bernd},
	journal={SIAM Journal on Matrix Analysis and Applications},
	volume={35},
	number={4},
	pages={1521--1542},
	year={2014},
	publisher={SIAM},
	doi = {10.1137/13094520X},
}

@article{gillard:2023,
	TITLE = {{Hankel low-rank approximation and completion in time series analysis and forecasting: a brief review}},
	AUTHOR = {Gillard, Jonathan and Usevich, Konstantin},
	URL = {https://hal.science/hal-03690201},
	JOURNAL = {{Statistics and Its Interface}},
	HAL_LOCAL_REFERENCE = {BioSiS},
	PUBLISHER = {{International Press}},
	VOLUME = {16},
	NUMBER = {2},
	PAGES = {287-303},
	YEAR = {2023},
	MONTH = Apr,
	DOI = {10.4310/22-SII735},
	HAL_ID = {hal-03690201},
	HAL_VERSION = {v1},
}

@article{CULLUM1981329,
	title = "{Computing eigenvalues of very large symmetric matrices—An implementation of a Lanczos algorithm with no reorthogonalization}",
	journal = {Journal of Computational Physics},
	volume = {44},
	number = {2},
	pages = {329-358},
	year = {1981},
	issn = {0021-9991},
	doi = {https://doi.org/10.1016/0021-9991(81)90056-5},
	url = {https://www.sciencedirect.com/science/article/pii/0021999181900565},
	author = {Jane Cullum and Ralph A Willoughby},
}

@article{Rodekamp:2024ixu,
	author = {Rodekamp, Marcel and Berkowitz, Evan and G\"antgen, Christoph and Krieg, Stefan and Luu, Thomas and Ostmeyer, Johann and Pederiva, Giovanni},
	title = "{Single Particle Spectrum of Doped $\mathrm{C}_{20}\mathrm{H}_{12}$-Perylene}",
	eprint = "2406.06711",
	archivePrefix = "arXiv",
	primaryClass = "cond-mat.str-el",
	doi = "10.1140/epjb/s10051-024-00859-1",
	journal = "Eur. Phys. J. B",
	volume = "98",
	pages = "36",
	month = "2",
	year = "2025",
}

@article{Zvonarev:2017,
	author = {Zvonarev, Nikita and Golyandina, Nina},
	year = {2017},
	month = {01},
	pages = {5-18},
	title = {Iterative algorithms for weighted and unweighted finite-rank time-series approximations},
	volume = {10},
	journal = {Statistics and its interface},
	doi = {10.4310/SII.2017.v10.n1.a1}
}

@article{Aubin:2010jc,
	author         = "Aubin, C. and Orginos, K.",
	title          = "{A new approach for Delta form factors}",
	booktitle      = "{Proceedings, 12th International Conference on
	Meson-nucleon physics and the structure of the nucleon
	(MENU 2000): Williamsburg, USA, May 31-June 4, 2010}",
	journal        = "AIP Conf. Proc.",
	volume         = "1374",
	year           = "2011",
	number         = "1",
	pages          = "621-624",
	doi            = "10.1063/1.3647217",
	eprint         = "1010.0202",
	archivePrefix  = "arXiv",
	primaryClass   = "hep-lat",
	reportNumber   = "JLAB-THY-10-1317",
	SLACcitation   = "%%CITATION = ARXIV:1010.0202;%%"
}

@article{Schiel:2015kwa,
	author         = "Schiel, Rainer W.",
	title          = "{Expanding the Interpolator Basis in the Variational
	Method to Explicitly Account for Backward Running States}",
	journal        = "Phys. Rev.",
	volume         = "D92",
	year           = "2015",
	number         = "3",
	pages          = "034512",
	doi            = "10.1103/PhysRevD.92.034512",
	eprint         = "1503.02588",
	archivePrefix  = "arXiv",
	primaryClass   = "hep-lat",
	SLACcitation   = "%%CITATION = ARXIV:1503.02588;%%"
}

\end{document}